\input psfig
\input mn.tex


\def\Msun{{\rm\,M_\odot}}

\def\kpc{{\rm\,kpc}}

\def\R1R2{{\rm R}_1^{'}{\rm R}_2^{'}}
\def\vecj{{\bf J}}

\def\vecw{{\bf w}}
\def\Hp{H_{\rm p}}
\def\avham{\langle H \rangle}
\def\avpot{\langle \psi \rangle}
\def\Mnep{M_{\rm N}}
\def\rnep{r_{\rm N}}
\def\Pa{{\rm P}_1}
\def\Ra{{\rm R}_1}
\def\Pb{{\rm P}_2}
\def\Rb{{\rm R}_2}

\def\Ba{{\rm B}_1}
\def\Bb{{\rm B}_2}
\def\Bc{{\rm B}_3}
\def\Jr{J_{\rm r}}
\def\Je{J_{\rm X}}
\def\Jc{J_{\rm circ}}
\def\Hsep{H_{\rm sep}}
\def\acos{{\rm \, acos}}
\def\asin{{\rm \, asin}}
\def\Ss{{\rm SS}^\prime}
\def\Op{\Omega_{\rm p}}
\def\Js{J_{\rm s}}
\def\Jf{J_{\rm f}}
\def\ws{w_{\rm s}}
\def\wf{w_{\rm f}}
\def\fr#1#2{\textstyle {#1\over #2}\displaystyle}

\def\piby2{{\pi \over 2}}

\def\t{\tilde}
\def\Jt{{\tilde J}}
\def\wt{{\tilde w}}
%
%
\def\spose#1{\hbox to 0pt{#1\hss}}
\def\lta{\mathrel{\spose{\lower 3pt\hbox{$\sim$}}
    \raise 2.0pt\hbox{$<$}}}
\def\gta{\mathrel{\spose{\lower 3pt\hbox{$\sim$}}
    \raise 2.0pt\hbox{$>$}}}
\def\today{\ifcase\month\or
 January\or February\or March\or April\or May\or June\or
 July\or August\or September\or October\or November\or December\fi
 \space\number\day, \number\year}
\newdimen\hssize
\hssize=8.4truecm  
\newdimen\hdsize
\hdsize=14.5truecm


\newcount\eqnumber
\eqnumber=1
\def\chaphead{}
 
\def\new{\hbox{(\rm\chaphead\the\eqnumber)}\global\advance\eqnumber by 1}
 
\def\first{\hbox{(\rm\chaphead\the\eqnumber a)}\global\advance\eqnumber by 1}
\def\last#1{\advance\eqnumber by -1 \hbox{(\rm\chaphead\the\eqnumber#1)}\advance
     \eqnumber by 1}
 
\def\ref#1{\advance\eqnumber by -#1 \chaphead\the\eqnumber
     \advance\eqnumber by #1}
 
\def\nref#1{\advance\eqnumber by -#1 \chaphead\the\eqnumber
     \advance\eqnumber by #1}

\def\eqnam#1{\xdef#1{\chaphead\the\eqnumber}}
 
 

\pageoffset{-0.85truecm}{-0.truecm}



\pagerange{}
\pubyear{1995}
\volume{000, 000--000}


\begintopmatter

\title{The Capture and Escape of Stars}

\author{J.L.\ Collett,$^1$ S.N.\ Dutta,$^{1,2}$ and N.W.\ Evans$^1$}

\affiliation{$^1$Theoretical Physics, Department of Physics, 1 Keble Road,
                 Oxford, OX1 3NP}
\vskip0.15truecm
\affiliation{$^2$Department of Physics and Astronomy, Ohio University,
Athens, OH 45701-2979, USA} 

\shortauthor{J.L.\ Collett, S.N.\ Dutta and N.W.\ Evans} 

\shorttitle{Capture and Escape} 


\abstract{The shape of galaxies depends on their orbital populations.
These populations change through capture into and escape from
resonance.  Capture problems fall into distinct cases depending upon
the shape of the potential well. To visualise the effective potential
well for orbital capture, a diagrammatic approach to the resonant
perturbation theory of Born is presented. These diagrams we call
equiaction sections. To illustrate their use, we present examples
drawn from both galactic and Solar System dynamics. The probability 
of capture for generic shapes of the potential well is calculated.

A number of predictions are made. First, there are barred galaxies
that possess two outer rings of gas and stars (type $\R1R2$). We show
how to relate changes in the pattern speed and amplitude of the bar to
the strength of the two rings. Secondly, under certain conditions,
small disturbances can lead to dramatic changes in orbital shape. This
can be exploited as a mechanism to pump counter-rotating stars and gas
into the nuclei of disk galaxies. Tidal resonant forcing of highly
inclined orbits around a central mass causes a substantial increase in
the likelihood of collision. Thirdly, the angular momentum of a
potential well is changed by the passage of stars across or capture
into the well. This can lead to the creation of holes, notches and
high velocity tails in the stellar distribution function, whose form
we explicitly calculate.}


\maketitle  


\section{1 Introduction}

Stars within galaxies belong to orbital families. The size and shape
of a galaxy determines the relative populations of these families.  As
a galaxy evolves, capture and escape of stars between these families
takes place. So, capture and escape are generic processes that will
have occurred many times in the history of galaxies.

Each orbital family has a parent periodic orbit -- that is, an orbit
that describes a closed figure. For example, in a spherical galaxy,
all stars belong to the family of tube orbits whose members librate
around the closed circular orbits.  An oval distortion in the centre
of this galaxy is supported by the family of box orbits, whose parent
periodic orbits are the radial orbits. As the distortion grows, stars
are transferred from the loop to the box family. This capture process
is important in the formation, maintenance and secular evolution of
non-axisymmetric structures, such as bars, rings and spiral arms
(e.g., Lynden-Bell 1973; Kalnajs 1973; Tremaine \& Weinberg 1984). 
Whether or not a trapped star remains trapped may depend on the
presence of a central black hole or mass concentration.  Close passage
can scatter a star away from its original orbit and thus cause a
gradual disruption of the population of orbits.

Let us turn to some specific problems. First, some barred galaxies,
such as NGC 5701, possess {\it two} outer rings of gas and stars. Buta
(1986), who labels these galaxies $\R1R2$, suggests that they may be
comprised of stars on periodic orbits aligned and anti-aligned with
the bar. As the bar evolves, is it possible for stars to be exchanged
between the rings?  Second, counter-rotating gas is present in some S0
and spiral galaxies (Bertola, Buson \& Zeilinger 1992).  For example,
the \lq evil-eye' galaxy NGC 4826 has an outer HI ring, which is
counter-rotating, whilst its interior gas is co-rotating (Braun,
Walterbos \& Kennicutt 1992). Recently, substantial counter-rotating
gas has also been reported in the spiral NGC 3626 (Ciri, Bettoni \&
Galletta 1995). A counter-rotating component is perhaps unsurprising
if the galaxy has suffered a retrograde merger or has undergone
substantial secondary accretion or infall. We shall show that
counter-rotating gas and stars are susceptible to large-scale orbital
shape changes under resonant perturbation. This provides a mechanism
for feeding gas into the centres of galactic nuclei. In a classic
paper, Tremaine \& Weinberg (1984) consider the origin of frictional
torques in stellar systems, showing the crucial r\^ole played by the
transfer of stars across the resonances. The shape of the effective
potential well at the resonance determines the dynamics of this
process. This in turn is very sensitive to orbital shape. We consider
cases in which this can lead to marked changes in the stellar
distribution function, including the formation of holes and notches.

\beginfigure*{1}
\centerline{\psfig{figure=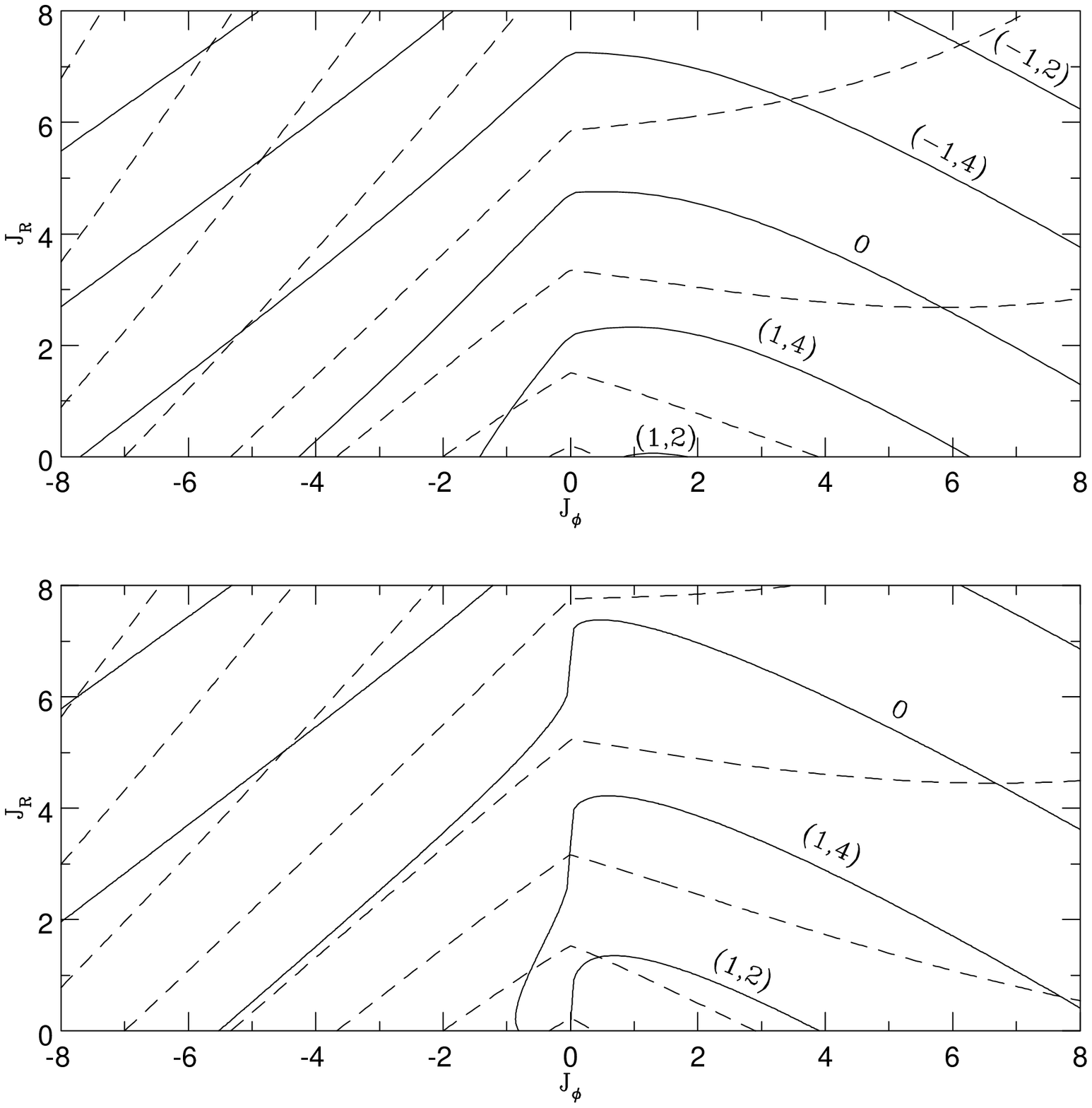,height=12truecm,width=12truecm}}
\caption{{\bf Figure 1.} The solid lines represent the reticulation 
of the $\vecj$-plane by ($\ell,m$) resonances defined by the closure
conditions [equations (2.1) and (2.6)]. The broken lines are the
allowed fast action paths. They intersect a particular resonance line
with the same gradient independent of the pattern speed. In the upper
panel, the background potential is an axisymmetric Binney disk
[equation (2.13)] and the pattern speed is taken as 0.1 (in units in
which $v_0 = R_c = 1$). The lower panel shows the effects of removing
the harmonic core and reducing the pattern speed. The background model
is an axisymmetric Mestel disk and the pattern speed is 0.075.}
\endfigure

\beginfigure{2}
\centerline{\psfig{figure=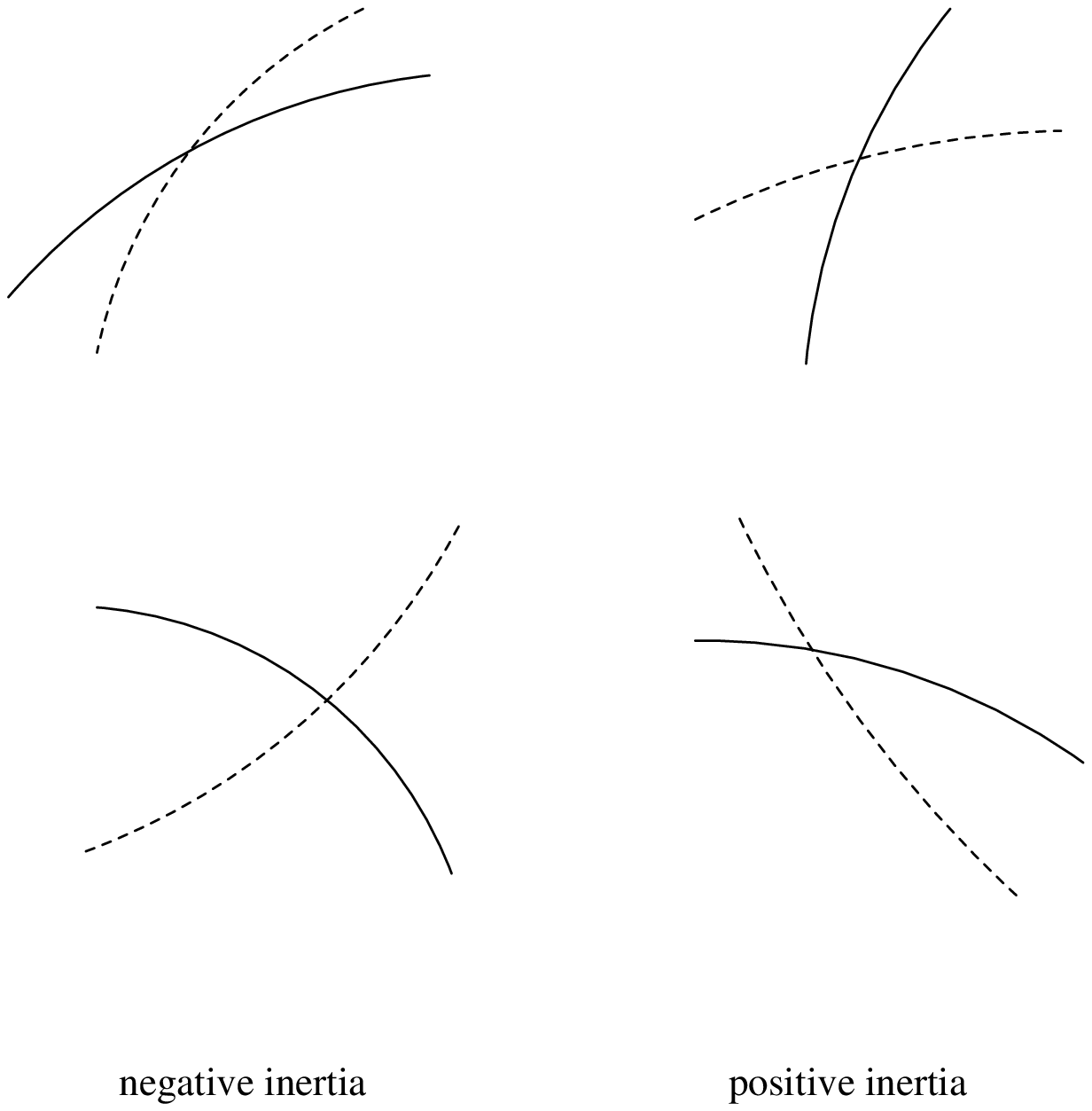,height=7.0truecm,width=7.0truecm}}
\caption{{\bf Figure 2.} The sign of the inertial response of the stellar 
orbits may be deduced from the intersection of the resonance lines and
the fast action tangent paths. The tangents to the broken lines give
the local evolutionary tracks.  In the left panels, $\Op$ decreases as
$\Js$ increases (negative inertia); in the right panels, $\Op$
increases as $\Js$ increases (positive inertia). It is evident on
comparison with Fig.~1 that regions of negative inertial response are
preponderant.  Positive inertial response, however, occurs for the
negative angular momentum continuation of the ($1,4$) resonance for
the Binney disk, as well as at the inner Lindblad resonance ($1,2$)
for the Mestel disk.  }
\endfigure

\noindent
The aim of our paper is to shed light on these matters. This is made
easier if we have a simple, physical picture of resonant escape and
capture, which we provide in Section 2.

\eqnumber =1
\def\chaphead{\hbox{2.}}
\section{2 The Equiaction Section} 

\noindent
Stellar orbits in a galactic disk nearly always form rosettes. 
Occasionally, the radial frequency $\kappa$ and the angular frequency
$\Omega$ of the star are commensurable, i.e., $\ell \kappa = m \Omega$
for some integers $\ell, m$. Then, the star's orbit is periodic and it
closes after $m$ radial librations and $\ell$ turns around the
centre. Even if the orbit is a rosette, it can be made to close by
moving to a rotating frame. To an observer rotating steadily in space
with angular velocity $\Op$, an orbit is closed if
\eqnam\resonant
$$\ell \kappa = m(\Omega - \Op) .\eqno\new$$
An orbit meeting this condition is {\it resonant}. Why are the
resonant orbits so important? Suppose a disturbance rotating at
angular frequency $\Op$ is applied to the disk.  On each traverse, the
resonant stars meet the crests and troughs of the perturbation
potential at the same spots in their orbits and this causes secular
change in the orbital elements. The non-resonant stars feel only
periodic fluctuations that average to zero. As the strength of the
perturbation increases, stars near the locus of exact resonance are
captured into libration around the parent periodic orbit. So, the
neighbourhoods of the resonances are the regions of a galaxy where a
disturbance can produce long term effects by changing populations of
orbital families.

Near a resonance, the star's motion can be nicely decoupled into two
disparate timescales -- {\it fast} and {\it slow} oscillations (e.g.,
Born 1927; Lynden-Bell 1973; Tremaine \& Weinberg 1984).
The orbital motion is then pictured as the fast traversal of a closed
figure together with the slow libration of its line of apsides.  Let
($R,\phi$) be polar coordinates in an axisymmetric galactic disk.
Associated with the two periodic motions in radius and azimuth on our
rosette orbit are actions ($J_R, J_\phi$) and angles ($w_R, w_\phi$)
(see e.g., Born 1927; Arnold 1978). One useful property of the
actions is their adiabatic invariance under slow dynamical change.
The Hamiltonian $H_0$ of stars in the axisymmetric disk is a function
of the actions alone. The frequencies $\kappa$ and $\Omega$ are simply
given by
$$\kappa = {\partial H_0 \over \partial J_R},\qquad
  \Omega = {\partial H_0 \over \partial J_\phi}. \eqno\new$$
Near a resonance -- where a combination of these frequencies is close
to the forcing frequency (2.1) -- the adiabatic invariance of ($J_R,
J_\phi$) breaks down. Nevertheless, there is a linear combination of
actions, specific to each resonance, that is preserved, namely the
circulation $\oint {\bf p}d{\bf q}$ around the closed figure. This is
the fast action.  In order to exploit this invariant, we perform at
each resonance a separate canonical transformation to the
corresponding slow and fast actions ($\Js,\Jf$) and angles ($\ws,
\wf$). This is effected by the generating function $S(\Js,\Jf;
w_R,w_\phi,t)$ (e.g., Lynden-Bell 1973; Tremaine \& Weinberg 1984;
Earn 1993)
$$S({\bf J}, {\bf w}) = \Js \Bigl[ w_\phi - {\ell\over m} w_R - \int
\Op dt \Bigr] + w_R \Jf,\eqno\new$$
so that
\eqnam\fastslowact
$$\eqalign{ &{\partial S\over \partial w_\phi} = J_\phi = \Js,\cr
            &{\partial S\over \partial \Js} = \ws = w_\phi - {\ell\over
             m}w_R - \int \Op dt,\cr
            &{\partial S\over \partial w_R} =J_R = \Jf - {\ell\over
             m}J_\phi, \cr
            &{\partial S\over \partial \Jf} =\wf = w_R,\cr
            &{\partial S \over \partial t} = -\Op \Js.\cr}\eqno\new$$
Therefore, the fast and slow actions are
\eqnam\fastslowacttwo
$$\eqalign{ \Jf &= J_R + {\ell \over m} J_\phi,\cr
            \Js & = J_\phi.\cr}\eqno\new$$
When the angular momentum changes sign, the resonance condition is
altered (Kalnajs 1977)
$$(m - \ell ) \kappa = m(\Omega + \Op),\eqno\new$$
where all the frequencies are taken as positive. The functional form 
of the fast and slow actions now becomes
\eqnam\negativebranch
$$\eqalign{ \Jf &= J_R - {\ell \over m} | J_\phi | + | J_\phi |,\cr
            \Js &= J_\phi.\cr}\eqno\new$$
The fast action is continuous at $J_\phi = 0$, even though its
functional dependence on $\vecj$ has changed. Kalnajs (1977) gives a
beautiful illustration of this point in terms of the zero angular
momentum orbits which are the common limits of the ($\ell,m$) and
($-\ell-m,m$) orbits. 

\beginfigure*{3}

\centerline{\psfig{figure=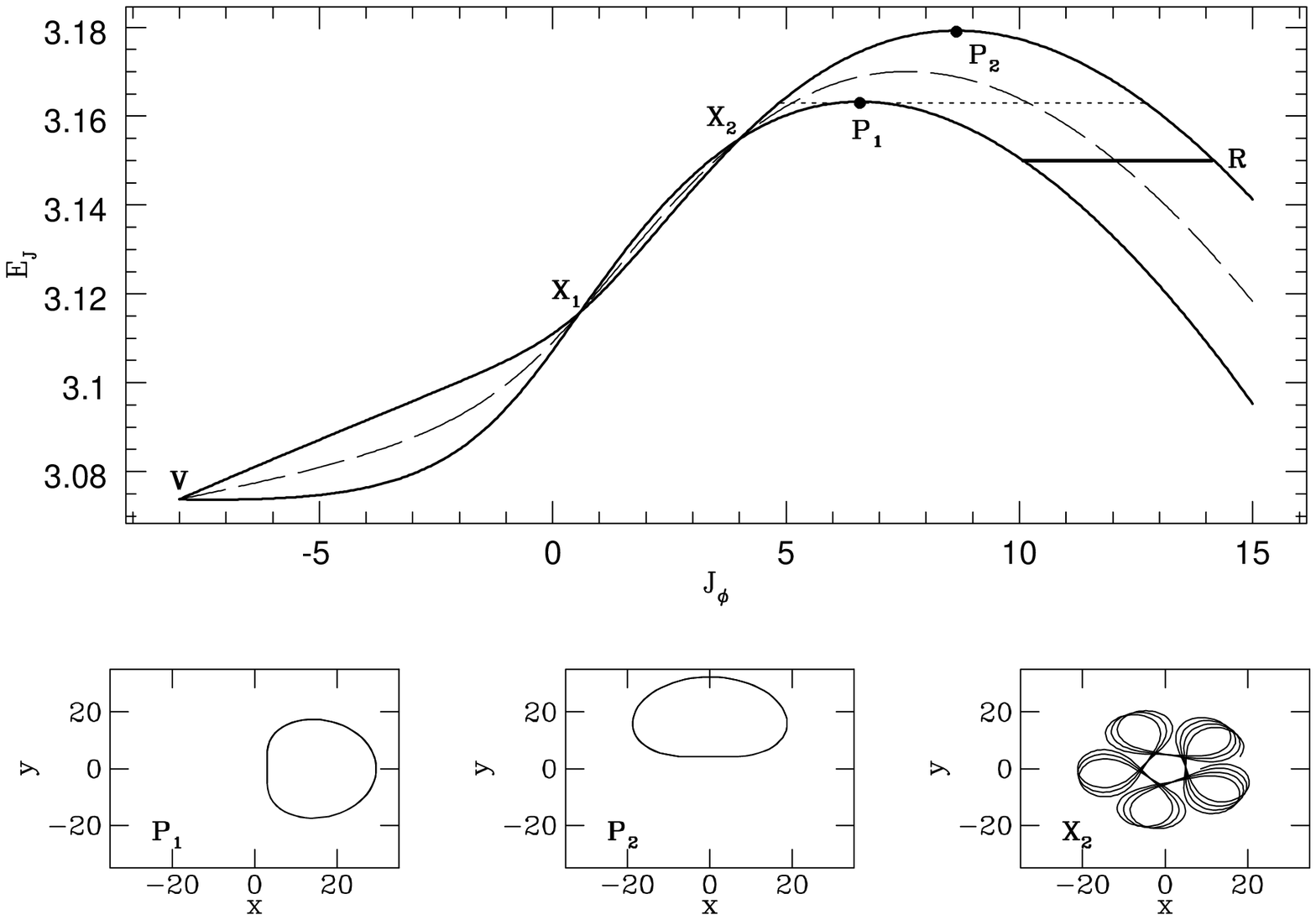,height=12truecm,width=10.875truecm}}

\caption{{\bf Figure 3.} An equiaction section at co-rotation in the
cored logarithmic model. The central seam (drawn in a broken
line) represents the unperturbed Hamiltonian in the rotating frame. The 
upper and lower boundaries represent the full values of the Hamiltonian 
at the crests ($\ws = \pi/2$) and the troughs ($\ws = 0$). The trajectory 
of every star is a horizontal line on the section. Above the separatrix 
(dotted line), the orbits are caught into libration. Below they rotate
and an example of a retrograde rotator is marked R. The periodic orbits 
$\Pa$ and $\Pb$ are recovered at the horizontal tangents. The planforms
of the orbits are shown below the section. (The model used is
(2.13) with $q =0.85$, $\Op = 0.05$, and $J_{f}= 8$ in units with 
$v_0 = R_c =1$). The bar axis is horizontal.}
\endfigure

\beginfigure{4}
\centerline{\psfig{figure=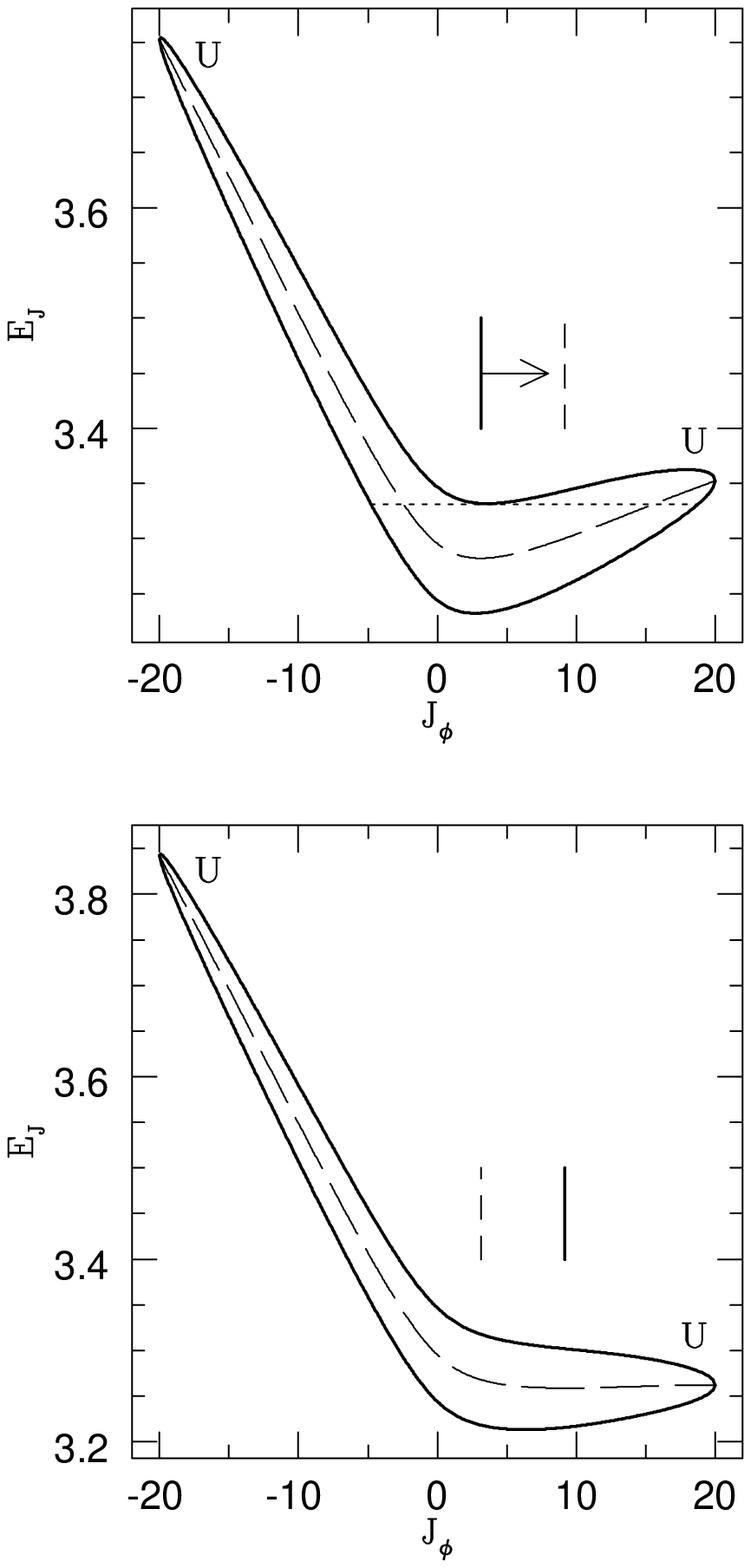,height=12truecm,width=\hssize}}
\caption{{\bf Figure 4.} When the pattern speed changes, the position of
exact resonance moves along the section. The first panel shows the
equiaction section for the Binney disc with $q=0.9, \Op = 0.01$
and $\Jf = 10$. The second panel is the section taken at the same value 
of the fast action, but with the higher pattern speed $\Op =
0.0145$. In this example, the section becomes flatter as the pattern
speed is increased. This makes the box orbits fatter. 
}
\endfigure

The reticulation of the $\vecj$-plane by the lines of exact ($\ell,m$)
resonance is shown in Fig.~1. The principal resonances are marked.
The tangents to the broken track correspond to allowed dynamical paths
because they conserve the local value of the fast action as prescribed
by equation (2.5). A star trapped at a resonance oscillates along this
tangent.  The set of all these line elements crossing a resonance
constitute a band of trapped stars, The shape and width of this band
depends on the amplitude of the perturbation and the figure of the
orbit. Together they determine an effective potential well in which
the oscillation takes place. To construct this well, we examine the
Hamiltonian of the orbits in the rotating frame.
\eqnam\hamiltonian
$$\eqalign{H =& H_0(\vecj) + {\partial S \over \partial t} + 
\Hp(\vecj ,\vecw, t)\cr
 =& H_0(\vecj)-\Op \Js +\Hp(\vecj,\vecw,t).
\cr}\eqno\new$$ 
where $\Hp$ represents the perturbation. Averaging this Hamiltonian
over the fast motion leads to the following equations of motion
\eqnam\aveqnsofm
$$\eqalign{{d \Js \over dt} =& -{\partial \avham \over \partial \ws} =
                      {\partial \avpot \over \partial \ws},\cr
{d \ws \over dt} =& {\partial \avham \over \partial \Js} =
                  \Omega (\vecj) - {\ell \over m} \kappa (\vecj) - 
                  \Op - {\partial \avpot \over \partial \Js},\cr}
                  \eqno\new$$
where the perturbation potential $\psi$ can be expanded as a Fourier 
series over the harmonics
\eqnam\fourierexp
$$\avpot = \sum_{m = -\infty}^{\infty} A_m (\vecj, t) \exp (im\ws).
\eqno\new$$
The angled brackets in equations (\aveqnsofm) and (\fourierexp) denote
averages over the fast phase.  If the perturbation does not depend
explicitly on time, then the Hamiltonian in the rotating frame or
Jacobi integral $E_{\rm J} =\langle H \rangle$ is conserved. A rough
idea of the dynamics is obtained by restricting ourselves to a single
harmonic component and expanding about the exact resonance to obtain
the approximate equations of motion
\eqnam\oneharmonic
$$\eqalign{ {\dot \Js} &= -mA_m(\vecj) \sin m\ws,\cr}\eqno\first$$
$$\eqalign{ {\dot \ws} & = D_1(\Js -\Jr) - {\partial A_m\over
            \partial\Js} \cos m \ws.\cr}\eqno\last b$$
The slow angle $\ws$ measures the inclination of the line of apsides
of the figure in the trough of the well. Equation (2.11a) illustrates
the angular momentum transfer when orbit and well are offset. Equation
(2.11b) contains the inertial response of the orbit $D_1$ defined as
\eqnam\inertialresponse
$$D_1 = {\partial^2 H_0 \over \partial \Js^2} = {\partial \Op \over
\partial \Js}.\eqno\new$$
This is analogous to the reciprocal of the moment of inertia of a
rigid body. Near-resonant orbits have the curious property that their
angular inertia depends on the forcing frequency.  The inertial
response of an orbit is positive when its angular velocity is
increased by an applied torque {\it as the fast action is held fixed.}
Any orbit will have a different inertial response at different pattern
frequencies because the fast action changes.  Let us remark that we
are following the notation of Tremaine \& Weinberg (1984) in writing
$D_n$ as the $n$th derivative of $\Op$ with respect to $\Js$. Earn \&
Lynden-Bell (1996; see also Earn 1993) refer to $D_1$ as the
cooperation parameter in their studies of disc models which have
regions of both positive and negative inertial response.

Fig.~2 shows how the inertial response of orbits can be deduced from
the intersections of the fast action tangents with the resonance
lines. When the resonance line and fast action tangent are nearly
parallel, the orbit possesses very large inertia and is able to move
some distance along the tangent without moving far from exact
resonance. The gradient term in (2.11b) will be of particular
importance when we investigate the capture and release of resonant
orbits in section 3.

We shall examine the trajectories of equations (2.9) in a slice
through phase space at constant fast action. These surfaces we call
equiaction sections (Evans \& Collett 1994; Collett 1995). The
effective potential well takes different forms in different sections.
We now consider specific examples of such sections at the three
principal resonances -- the co-rotation ($0,2$) and the inner and
outer Lindblad resonances ($\pm 1,2$). Low order resonances are the
most important because the resonant orbits display the most marked
deviations from axisymmetry and therefore couple most strongly to
simple non-axisymmetric patterns.

The equiaction sections can be compared to Poincar\'e surfaces of
section (see e.g., Gutzwiller 1990). These have been used extensively,
for instance, by Contopoulos and collaborators to map the orbital
structure of bar-like potentials (e.g., Contopoulos \& Papayannopoulos
1980). Our interest here is in time-dependent problems and the
equiaction sections have the advantage that the changes in orbital
families are more clearly depicted. Donner (1979) drew diagrams
reflecting the pendulum-like solutions of the equations (\aveqnsofm)
very close to the exact resonance. These are related to equiaction
sections, but the trajectories of stars were parabolae rather than the
straight lines here.

\beginfigure*{5}

\centerline{\psfig{figure=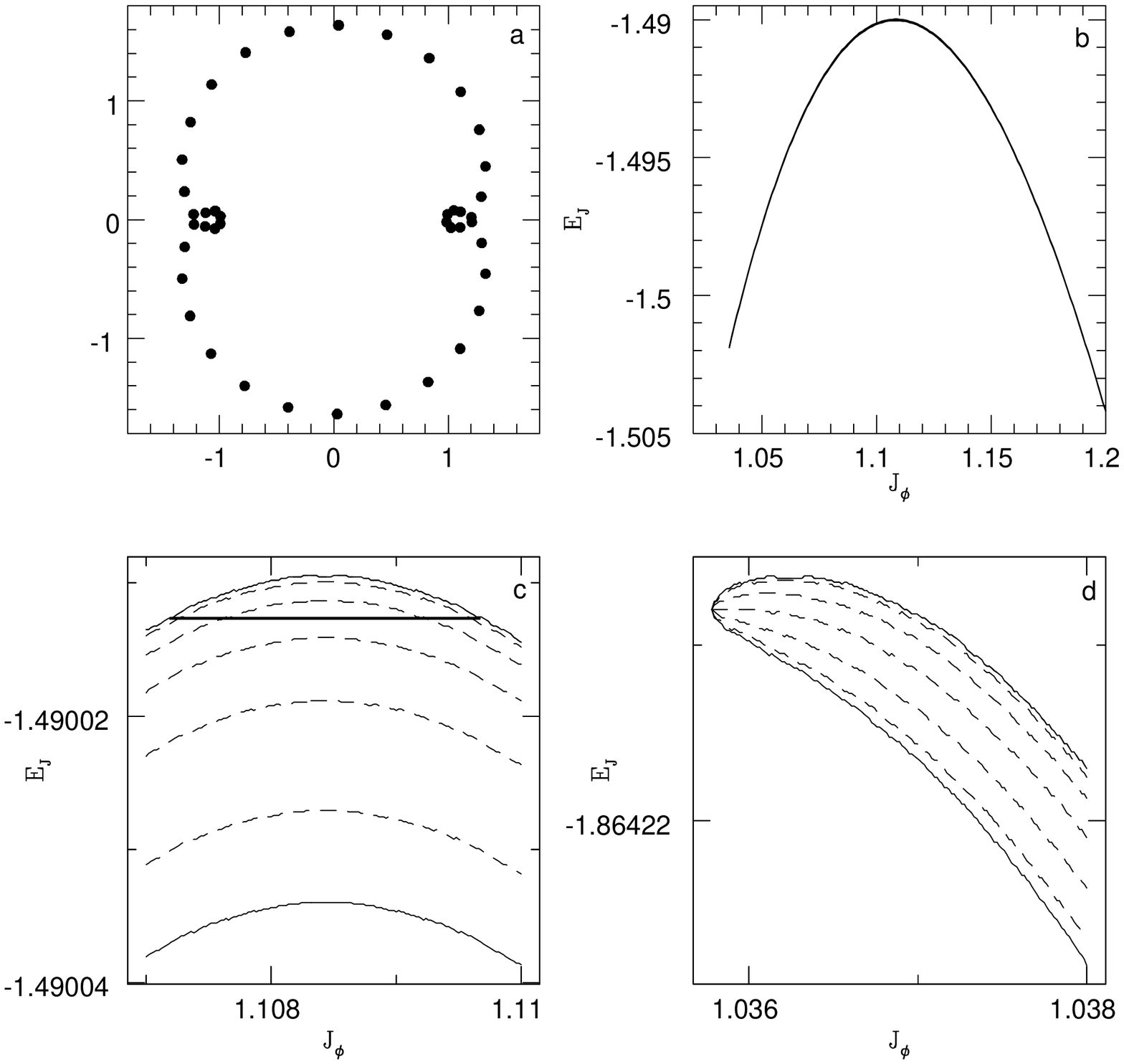,height=12truecm,width=12truecm}}

\caption{{\bf Figure 5.} (a) The closed orbit of Pluto in the 
frame co-rotating with the mean motion of Neptune is shown. The time
interval between the plotted points is $\sim 4,500$ days. The axis of
the line of apsides librates with an amplitude of $\sim 38^\circ$.
(b) The present-day equiaction section of Pluto. Pluto has worked its
way rightwards along the section as Neptune's orbit expanded. (c) A
detail of the resonant part of the section. The dashed lines show the
dependence of the energy on the orientation of the figure.  They
correspond to inclination angles of $15^\circ, 30^\circ, 45^\circ,
60^\circ$ and $75^\circ$ of the line of apsides. The influence of many
harmonics in the potential is evident in the dispersal of these
contours.  The present day orbit of Pluto is shown as a bold
horizontal line. (d) A detail of the tip of the section when the
radius of Neptune was $\sim 82 \%$ of its present value and Pluto was
trapped from a near-circular orbit. At the initial capture, the $m=2$
component in the perturbing potential was dominant, as indicated by
the symmetric placing of the dashed lines. [All the equiaction
sections are drawn with $\Jf = -0.518$ in units with $G=\Msun=\rnep
=1$].  }
\endfigure

\subsection{2.2 Co-rotation ($\ell = 0$) resonance}

At co-rotation, the resonant orbits are, for near-circular orbits,
Lindblad epicycles. Their small size means that they can feel a
significant orbit-averaged potential (because they sample the
perturbing potential in a small patch). This diminishes as the orbit
becomes larger, except when the perturbations has the same symmetry as
the orbit. The conserved fast action is the radial action [see
equation (\fastslowacttwo)] -- in Fig.~1, the fast action tangents
that intersect the ($\ell = 0$) resonance line are horizontal. When we
draw an equiaction section, we extend this tangent and then make our
cut. When the cut reaches $J_\phi = 0$, it changes direction in accord
with the new fast action (2.7).

In our dynamical model, the underlying Hamiltonian $H_0$ is that of
the axisymmetric cored Mestel (1963) disk. The rotating disturbance,
$H_p$, is a pure $m=2$ component, so that
\eqnam\binneydisk
$$\eqalign{H =& H_0 + H_p = \fr12 p_R^2 + \fr12 {p_\phi^2 \over R^2}
+ \fr12 \log ({\t R}_c^2 + R^2)\cr
&+ \epsilon {R^2\over {\t R}_c^2 + R^2}\cos 2(\phi - \int \Op dt )
.\cr}\eqno\new$$
When $\epsilon = (1 - q^{2})/(1 + q^2)$ and $R_c^2 = \fr12 {\t R}_c^2 
(1 + q^{-2})$, then the entire Hamiltonian replicates the weakly 
non-axisymmetric rotating Binney disk (see Binney 1982; Binney \& 
Tremaine 1987, p. 126)
\eqnam\binneydisktrue
$$H = \fr12 p_R^2 + \fr12 {p_\phi^2 \over R^2} - \Op p_\phi
+ \fr12 \log( R_c^2 + x^2 + y^2 q^{-2}).\eqno\new$$
The density corresponding to the potential (\binneydisk) is positive
for $0 \le q \le 1$. Using this Hamiltonian, a typical equiaction
section is drawn at co-rotation in Fig.~3. The central seam is the
unperturbed Hamiltonian in the rotating frame, $H_0(\vecj) - \Op
\Js$. The curvature of the central seam has a simple physical
interpretation. When the curvature is downward, as it is rightward of
point ${\rm X}_1$ in Fig.~3, then the inertial response
(\inertialresponse) is negative. The envelope of the equiaction
section is drawn by finding the orbit-averaged perturbation potential,
$\avpot$. This provides the effective potential within an equiaction
section. (A simple mechanical model of this is provided by the jointed
arm of Appendix A). For each orbit, $\avpot$ is computed in the
particular rotating frame in which the orbit closes. Since orbits
conserve the Jacobi integral in the frame of the perturbation, they
are horizontal lines on an equiaction section, internal to and bounded
by the envelope.  The shape of the envelope governs the range of the
angular momentum exchange of each orbit with the perturbation. The
periodic orbits $\Pa$ and $\Pb$ are recovered at the horizontal
tangents and experience no angular momentum exchange. These orbits
bifurcate, or split, from the position of exact resonance, which
corresponds to the maximum of the central seam. The splitting of the
orbits is obtained from (\aveqnsofm) as
\eqnam\splittingeq
$$\Delta \Js = 2 {\partial \avpot\over\partial \Op}.\eqno\new$$
$\Pb$, though at a maximum of the effective potential, is stable in the 
sense that orbits close by librate about it. $\Pa$ is unstable for it 
will, if disturbed, either rotate if it loses energy or perform a 
large amplitude libration about $\Pa$ if it gains energy. The dotted 
horizontal line is a separatrix, above which the stars are trapped. 
Below the separatrix, the orbits are retrograde rotators to the right 
(an example of which is the horizontal line labelled R) and prograde 
rotators to the left.

The capture and scattering between orbital families depends critically
on the shape of the envelope close to the resonance. Particularly
interesting dynamics can occur when the resonance lies close to points
at which the envelope crosses or meets the seam (when the gradient
term in (\oneharmonic) becomes important). There are three cases of
special interest:

{\it
\noindent
(1) U points, where the envelope terminates on the seam at a point
corresponding to the circular orbit $\Jc$ and the amplitude $A_m$ is
proportional to $|J - \Jc|^{1/2}$,

\noindent
(2) V points, where the envelope terminates on the circular orbit,
but the amplitude $A_m$ is now proportional to $|J - \Jc|$,

\noindent
(3) X points, where the envelope crosses the seam and $A_m
\propto |J - \Jc|$.
}
 
In Fig.~3, we see three points where the orbit-averaged potential
vanishes. The leftmost point of the equiaction section is a V-point
and corresponds to a circular orbit.  Further along the equiaction
section, the orbit-averaged potential changes sign twice. These
crossing points are X-points. The occurrence and position of
X-points depends on the radial form of the perturbation potential
encountered by the eccentric orbits along the section. X-points are
particularly common for minor resonances where the orbital shape is
quickly changing. The dynamics close to an X-point, and its
consequences for angular momentum transport, are explored in Section
3.1.

\subsection{2.3 Inner $(1,2)$ and Outer $(-1,2$) Lindblad resonance}

Fig.~4 shows equiaction sections for an inner Lindblad resonance
(ILR). The form of the sections here is of particular interest
because, amongst the principal resonances, the inner Lindblad resonant
frequencies are the most slowly varying with radius in galaxies. The
curvature of the central seam may now be upward corresponding to
positive inertial response. The figure presents a pair of equiaction
sections showing the orbital families associated with a decelerating
bar. The trough of the well is in each case represented by the lower
boundary of the section, the crest of the well by the upper
boundary. As the pattern speed increases, the position of exact
resonance moves rightwards (as the inertial response is
positive). This is indicated by the arrow.  In the lower panel, the
end of the section is markedly flatter, showing that the angular
momentum and orbital shape varies substantially during a libration.

The envelope closes with a characteristic tip, labelled U on the
figure.  This is different from the V-tip found at
co-rotation. Although it is often the case in dynamics that librators
are divided from rotators by a separatrix, or oscillation of infinite
period, this does not necessarily happen close to these tips. The
dividing line between the rotators and librators runs through the
circular orbit. The escape from trapping occurs through the circular
orbit, on which the torque vanishes. An example of this class of
trapped motion is provided by the Galilean satellites, Io, Europa and
Ganymede.  Europa is at ILR with respect to Io and Ganymede at ILR
with respect to Europa. This configuration is maintained by the small
eccentricity stabilisation mechanism (e.g., Lynden-Bell \& Kalnajs
1972; Peale 1976; Goldreich \& Tremaine 1981; Binney \& Tremaine 1987,
p. 151), in which the eccentricity forced upon the near-circular orbit
leads to the ensnaring torque. A mechanism such as tidal torquing --
which circularises the periodic orbit by pushing it towards the end of
the tip -- weakens the coupling of the satellites.

An example of the class of trapped motion in which there is a
separatrix is provided by Neptune and Pluto (the large eccentricity
stabilisation mechanism of Peale (1976)). Pluto's eccentric orbit
forms an almost closed figure in the frame of Neptune's mean motion
(Cohen \& Hubbard 1965; see also Fig.~5(a)). Pluto lingers in \lq\lq
the ears" of the orbit and the net torque then tries to align the
major axis of the orbit with Neptune's position. This is an outer
Lindblad resonance ($\ell = -1, m = 2$). In order to draw an
equiaction section for this problem, the point mass perturbation from
Neptune must be time-averaged around the closed figure of Pluto. The
potential contains many more components than the single harmonics
considered above. The monopole merely contributes a constant to the
orbit-averaged potential, whereas the dipole vanishes if the centre of
mass does not move. Neptune is effectively replaced by two masses
$\fr12 \Mnep$ placed at the radius of Neptune's orbit $\rnep$ and
arranged fore and aft of the sun. Fig.~5 shows the full equiaction
section for Pluto and two expanded details.  Note that in the outer
Lindblad resonant sections, the curvature of the central seam is
downward, reflecting the negative inertial properties of the stars in
a Keplerian potential. The upper and lower boundaries of the section
now generally correspond to the trough and the crest of the potential
well, reversing the roles they had at the inner Lindblad
resonance. This is obvious from Fig.~5(a), as Pluto's averaged
potential is most negative when the long-axis of its orbit is aligned
with the trough of the well and so the \lq\lq ears" are aligned with
the crest.

Malhotra (1993) has suggested that Pluto was initially captured from a
near-circular orbit. As Neptune was driven outwards by planetisimal
expulsion, Pluto's orbit expanded but remained trapped. It moved
rightwards along the section and its eccentricity increased. Its
initial and present positions on the section are illustrated in the
details. They are calculated by taking the present day eccentricity of
Pluto as 0.25 (Allen 1973). In units in which $\rnep = G = \Msun =1$,
it follows that the conserved fast action of Pluto $\Jf =
-0.517$. Fig.~5(b) shows the complete equiaction section of Pluto. Its
thinness is in striking contrast to the earlier sections dealing with
galactic resonances. This graphically illustrates one of the important
differences between celestial mechanics and galactic dynamics.
Non-axisymmetric disturbances in the solar system are comparatively
feeble. Fig.~5(c) is an expanded detail of the resonant portion of
Fig.~5(b). The bold horizontal line shows the present-day orbit of
Pluto, showing the $38^\circ$ amplitude of libration of the line of
apsides (Cohen \& Hubbard 1965). The dashed lines show the dependence
of the averaged potential on the relative orientation of the figure of
Pluto and Neptune. In its present position, the influence of many
harmonics is clear in the asymmetrical dispersal of the contour
lines. If Pluto was captured from a circular orbit, then -- assuming
the constancy of the fast action -- this must have occurred when the
radius of Neptune's orbit was $24.6\, {\rm AU}$, or $82 \%$ of its
present value. Fig.~5(d) shows the point of capture.  Note that when
the resonance is near the circular orbit, the perturbation is
dominated by a single $m=2$ harmonic. This process is reversible, so
that were Neptune's orbit to contract, Pluto's eccentricity would be
reduced. This will be true, too, of galactic stars, as a massive black
hole slowly descends into the galactic centre. Stars on eccentric
orbits may be caught as the resonance moves to a region of greater
frequency and released at smaller eccentricities when the amplitude
contracts and the curvature of the section decreases.  Indeed, each
star may pass through a succession of resonances during this process,
so that its motion in the $\vecj$-plane approximates to a series of
small linear steps, parallel in each case to the appropriate fast
action.

\eqnumber =1
\def\chaphead{\hbox{3.}}
\section{3 Capture and Escape}

\beginfigure*{6}

\centerline{\psfig{figure=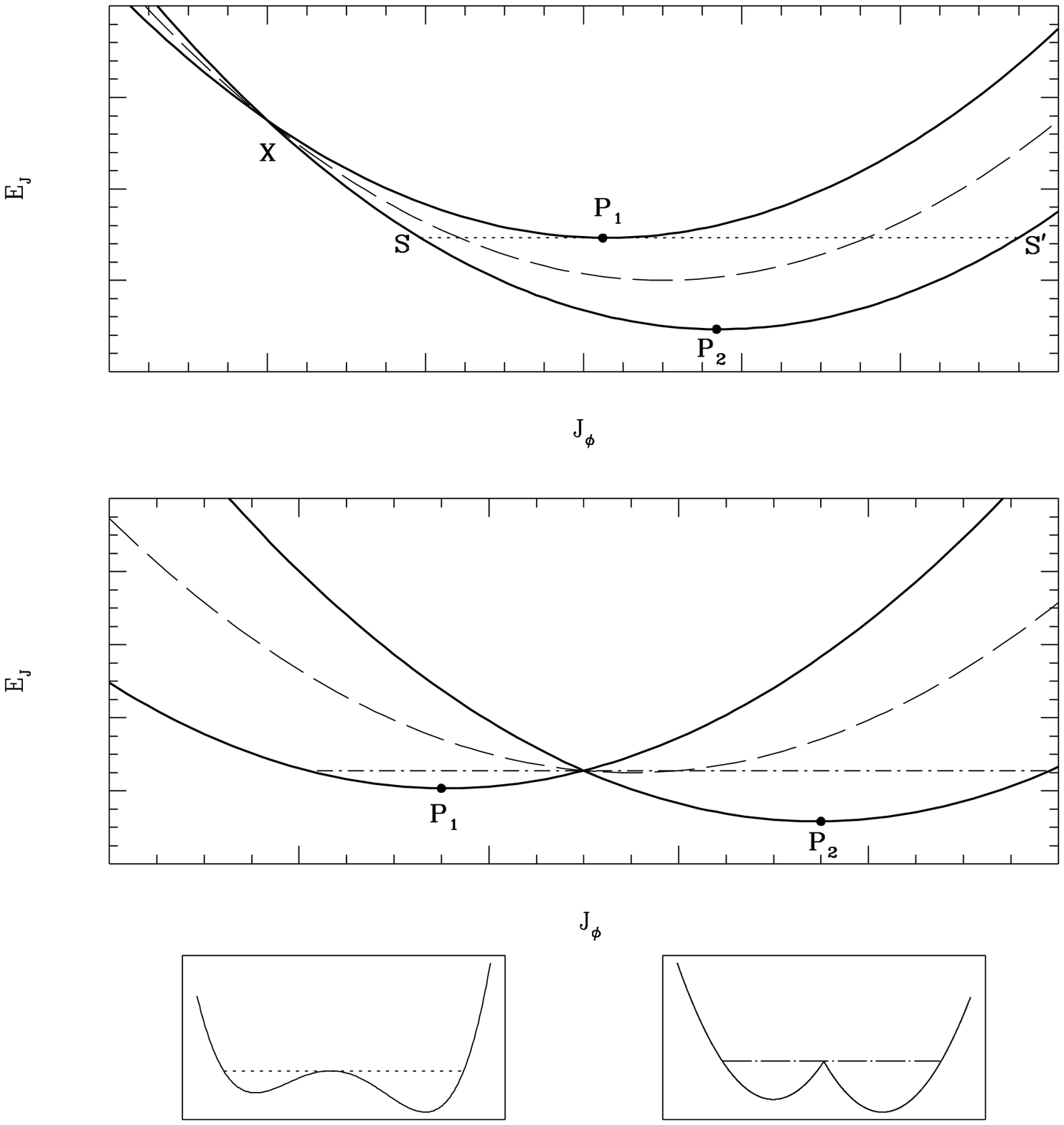,height=12truecm,width=10.875truecm}}

\caption{{\bf Figure 6.} The upper panel shows the generic features
of an equiaction section when a resonance lies close to an X-point.
The periodic orbits $\Pa$ and $\Pb$ are equally spaced on either side
of exact resonance. The separatrix $\Ss$ divides the librating family
associated with $\Pb$ from the two sets of rotators -- in this case,
prograde above $\Pa {\rm S}^\prime$ and retrograde above ${\rm
S}\Pa$. As the pattern speed is decreased, the position of exact
resonance moves towards the X-point. A critical point is attained when
$\Pa$ reaches the X-point. Beyond this, the equiaction section takes
the form shown in the lower panel with both $\Pa$ and $\Pb$ siring
families of librating orbits. Above the dot-dashed line, the orbits
are still rotators. Note that this line divides librators from rotators,
but does not correspond to an orbit of infinite period. It is a
pseudo-separatrix. The difference between a separatrix and a 
pseudo-separatrix is illustrated in the two small boxes. They show
one-dimensional motion in two different double potential wells. When
a cusp separates the wells, there is a pseudo-separatrix.
}
\endfigure

\noindent
Suppose Neptune's orbit continues to expand. Will Pluto remain trapped
forever? Provided the evolution remains adiabatic, it is the shape of
the section that describes the ultimate fate of Pluto. Two changes
occur in the section. First, the position of exact resonance moves
outwards and the curvature of the section is correspondingly
decreased. Second, the restoring torque diminishes and the envelope
constricts. Both effects lead to a slow increase in the amplitude of
libration and eventually Pluto may escape from Neptune's grasp.

The issues of escape and capture are the subject of our attention
here. The investigation of resonant capture in astronomy begins with
Goldreich's (1965) explanation of the occurrence and maintenance of
commensurable motions in the solar system. Subsequently, Yoder (1973,
1979) gave a diagrammatic method of calculating capture probabilities
for pendulum-like systems, while Henrard (1982) devised a general
treatment for any problem with one degree of freedom based upon the
use of adiabatic invariants. A comprehensive summary of the classical
work is contained in Henrard (1993).

Suppose a potential well of fixed amplitude has a slowly changing
pattern speed. As each star along the section encounters the well, it
may be captured or pass across the well and escape into
counter-rotation. Our ignorance of the exact phase at which the star
encounters the well's separatrix means that we must allocate
probabilities to these two motions. These probabilities reflect the
relative growth of the phase areas associated with the two modes of
motion. This can be deduced from the shape of the separatrix, which in
turn is set by the form of the section. We can distinguish three
regimes which need separate treatment. These are (i) in the vicinity
of an X-point or V-point, (ii) near a U-point and (iii) in a region of
slow variation in the amplitude of the envelope (e.g., such as close
to $\Pa$ in Fig.~3).  Of course, capture probabilities have been
calculated before in the context of celestial mechanics (e.g., Henrard
\& Lema\^itre 1983; Borderies \& Goldreich 1984). Here, we shall present
modified and generalised forms of these expressions, appropriate to
galactic resonances.

\subsection{3.1 Capture Probabilities near X- and V-points}

The Hamiltonian close to an X-Point may be written
\eqnam\xpointham 
$$H(J, w) = H_0 (J, w) + A_2 (J - \Je) \cos 2 w.
\eqno\new$$
Here, $\Je$ is the critical action corresponding to the X-point and
$A_2$ is a constant describing the amplitude of the well.  Although
(\xpointham) corresponds to the special case of the $m = 2$ harmonic,
our method of derivation is general. When the position of exact
resonance lies close to the X-point, the Hamiltonian may be Taylor
expanded to give
\eqnam\expandham
$$H(J,w) = \fr12 D_1 (J - \Jr)^2 + A_2 (J-\Je)
\cos 2w .\eqno\new$$
Here, $\Jr$ is the point of exact resonance and we assume that $D_1$
does not vanish, although, of course, it may be either positive or
negative depending on the sign of the inertial response. Now, we
linearly translate the action $J \rightarrow J - \Je$, so that the
Hamiltonian has the canonical form (dropping unimportant additive
constants)
\eqnam\canonicalham
$$H(J,w) = \fr12 D_1 J^2 - D_1 J (\Jr - \Je) + A_2 J
\cos 2w.\eqno\new$$
There are three parameters in (\canonicalham), namely $D_1, 
A_2$ and the difference $\Jr -\Je$. As this generic problem is 
invariant under scaling, we introduce the dimensionless action 
${\tilde J}$
$${\tilde J} = \fr12 \biggl| D_1 {J \over A_2} \biggr| .\eqno\new$$
Our problem is now characterised by a single parameter $\lambda =
\fr12 | D_1 (\Jr - \Je)/ A_2 |$. (When $\Jr \approx \Je$, this is just
$\fr12 | \Delta \Omega /A_2 |$, where $\Delta \Omega$ is the offset
of the X-point resonant pattern speed). The scaled Hamiltonian is very
simple, namely:
\eqnam\scaledham
$${\tilde H} = {\tilde J}^2 - 2\lambda {\tilde J} +
               {\tilde J} \cos 2 w.\eqno\new$$
The periodic orbits or equilibria are the fixed points of Hamilton's
equations. The stable equilibrium has coordinates $({\tilde J} =
\lambda + \fr12, w = \fr{\pi}{2})$ and is marked $\Pb$ in the upper panel of
Fig.~6. The unstable equilibrium has coordinates $({\tilde J} =\lambda
- \fr12, w = 0$) and is marked $\Pa$. When the position of exact
resonance gets too close to the X-point, the unstable equilibrium
disappears. As it crosses, it becomes a second stable equilibrium
point $\Pa$ marked on the lower panel of Fig.~6. The behaviour at a
V-point is similar to that at an X-point. If the X-point is
stationary, it is an immutable barrier. Stars, however, can pass from
one side of an X-point to another if the X-point itself is in motion,
which can happen when the perturbation's radial shape changes.

To calculate capture probabilities, we must work out how the phase area
of the trapping region changes along the section (this is quoted as 
\lq Kruskal's theorem' in Cary, Escande \& Tennyson 1986). So, we need 
to evaluate the actions associated with the two branches of the
separatrix ${\rm S}\Pa$ and $\Pa {\rm S}^\prime$ marked on the upper
panel of Fig.~6. The separatrix is a level curve of the Hamiltonian
(\scaledham). The value of the Hamiltonian marking the separatrix is
(inserting the action-angle coordinates of $\Pa$)
\eqnam\separatrix
$$\Hsep = - \lambda^2 + \lambda - \fr{1}{4}.\eqno\new$$
The points ${\rm S}$ and ${\rm S}^\prime$  are the two solutions of 
(\scaledham) with $H = \Hsep$ and $w= \fr{\pi}{2}$. The actions 
corresponding to these points are 
\eqnam\bandbdash
$${\t J}_{\rm S} = \lambda + \fr12 - \sqrt{2\lambda},
\qquad {\t J}_{\rm S^\prime} = \lambda + \fr12 + \sqrt{2\lambda}.
\eqno\new$$
The equation of the separatrix is
$${\tilde J} (w) = \lambda  - \fr12 \cos 2 w \pm
                 [2\lambda \sin^2 w - \fr{1}{4} \sin^2 (2w) ]^{1/2},
                 \eqno\new$$
where the negative sign corresponds to the branch ${\rm S}\Pa$
(retrograde rotators when $D_1$ is positive as in Fig.~6, prograde
rotators when $D_1$ is negative) and the positive sign to the branch
$\Pa {\rm S}^\prime$ (prograde rotators when $D_1$ is positive,
retrograde rotators when $D_1$ is negative). It is now straightforward
to evaluate the actions of the trapped orbit and the prograde and
retrograde rotators near the separatrix of Fig.~6. This leads to the
phase areas enclosed by these orbits:
\eqnam\actionsoforbits
$$\eqalign{&S_{\rm ret} = \lambda \acos \Bigl( {1\over \sqrt{2
\lambda}}\Bigr)  -\fr12 (2\lambda -1)^{1/2}, \cr
&S_{\rm prog} = \pi \lambda - \lambda \acos \Bigl( {1\over \sqrt{2
\lambda}}\Bigr) + \fr12  (2\lambda -1)^{1/2}, \cr
&S_{\rm trap} = \pi \lambda - 2\lambda \acos \Bigl( {1\over \sqrt{2
\lambda}}\Bigr) + (2\lambda -1)^{1/2}.}\eqno\new$$
To calculate capture probabilities, we find out how these three
areas change under infinitesimal variations of the parameter
$\lambda$. They are constrained by the conservation of phase area, so
that here
$${d S_{\rm trap}\over d \lambda} = {d S_{\rm prog}\over d\lambda} 
- {d S_{\rm ret}\over d \lambda} =0.\eqno\new$$
The general expression for the capture probability is provided by
Henrard (1993) as:
\eqnam\capturehenrard
$$P ={\displaystyle  {dS_{\rm trap}\over d \lambda}
      \over \displaystyle  {dS_{\rm trap}\over d \lambda}
      + {dS_{\rm ret}\over d \lambda} }
      ={\displaystyle  {dS_{\rm prog} \over d \lambda} - 
       {d S_{\rm ret}\over d \lambda}
      \over  \displaystyle {d S_{\rm prog}\over d \lambda}}.\eqno\new$$
For the problem at hand, the probability of capturing a prograde 
rotator becomes
\eqnam\captureatx
$$P = {\displaystyle \pi - 2\acos \Bigl( {1\over \sqrt{2 \lambda}}\Bigr) 
      \over  \displaystyle \pi - \acos \Bigl( {1\over \sqrt{2 \lambda}}
      \Bigr)}
     = {\displaystyle 2 \over  \displaystyle 1 + {\pi\over 2 \asin (
     {1\over \sqrt{2 \lambda}}) }}. \eqno\new$$
Note that $1/2 < \lambda < \infty$. Using different methods and
within the context of celestial mechanics, this formula was derived
for capture at a V-point by Yoder (1973). We have shown the same
result is valid for capture at an X-point. The difference, however, is
that the periodic orbit $\Pa$ disappears at a V-point, but crosses and
sires a librating family at an X-point.
 
\beginfigure{7}

\centerline{\psfig{figure=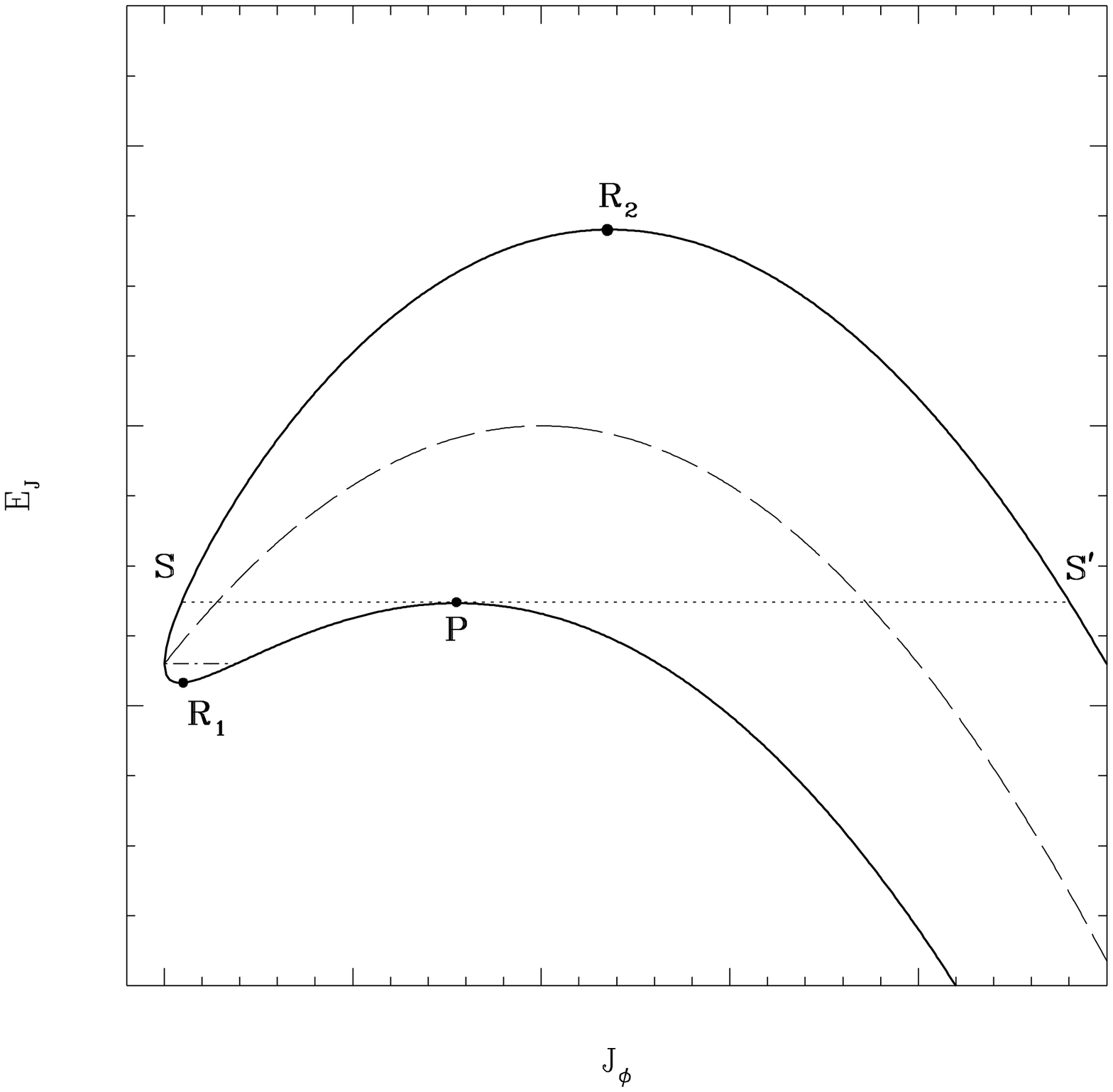,height=\hssize,width=\hssize}}

\caption{{\bf Figure 7.} The anti-aligned ($\Ra$) and aligned
($\Rb$) periodic orbits when a resonance lies close to a U-point.
These orbits build the two rings in $\R1R2$ galaxies. The unstable 
periodic orbit ${\rm P}$ divides the separatrix (drawn as a dotted
line) into two unequal branches ${\rm SP}$ and ${\rm PS}^\prime$. 
Above $\Ra$, there is a pseudo-separatrix --  shown as
a dot-dashed line. Although this divides librators from rotators, 
it does not correspond to an orbit of infinite period.} 
\endfigure

\subsection{3.2 Capture Probabilities near U-points : The Two 
Ring Problem}

A V-point is obtained, for example, when a $(-1,1)$ resonant orbit is
subjected to a pure $m=2$ harmonic. In a perturbation which shares the
lop-sided ($m=1$) symmetry of the orbit, the section possesses a
U-point. In the first case, $\avpot$ depends on the square of the
eccentricity $e^2$; in the second case, $\avpot \propto e$. This is
evident in Fig.~5(d), where the end of the section is dominated by the
$m=2$ component. Generally, we will find a U-point for any
near-circular resonant orbit in the presence of a perturbation with
the same symmetry.

As the methods of derivation are the same as in the previous section
and the result is well-known in celestial mechanics, we shall just
quickly sketch the theory before passing on to applications. The 
Hamiltonian in the rotating frame near the U-point is
$$H(J, w) = H_0 (J, w) + A_2 |J - \Jc|^{1/2} \cos 2w,
\eqno\new$$
where $\Jc$ is the action of the circular orbit at the very tip.
On Taylor expansion about the resonant action $\Jr$, this yields the 
Hamiltonian:
\eqnam\expandham
$$H(J,w) = \fr12 D_1 (J - \Jr)^2 + A_2 |J-\Jc|^{1/2}
\cos 2w.\eqno\new$$
By defining scaled action-angle coordinates, 
$$\Jt = 2^{1/3} \Bigl|{ D_1 \over A_2} \Bigr|^{2/3}| J - \Jc |,\qquad
\wt = w/2.\eqno\new$$
the Hamiltonian can be cast into Henrard's (1993) form
\eqnam\upointham
$${\t H} = {\t J}^2 - 2 \lambda {\t J} - 2 (2 {\t J})^{1/2} \cos {\t w},
\eqno\new$$
where 
\eqnam{\deflambda}
$$\lambda = 2^{1/3} \Bigl|{ D_1 \over A_2} \Bigr|^{2/3}
|\Jr -\Jc|.\eqno\new$$
It is useful to introduce the canonical coordinates
$$x= (2 {\t J})^{1/2} \cos {\t w}, \qquad y = (2 {\t J})^{1/2} \sin
{\t w},\eqno\new$$
so that the scaled Hamiltonian becomes
$${\t H} = \fr14 (x^2 + y^2)^2 - \lambda (x^2 + y^2) - 2x.\eqno\new$$
The equilibria satisfy
$$x^3 - 2\lambda x - 2 = 0, \qquad y = 0.\eqno\new$$
The three roots of the cubic correspond to two stable and one unstable
equilibria. The coordinates of the unstable fixed point are:
$$x_u = - \Bigl( {2\lambda \over 3} \Bigr)^{1/2} ( \cos \Delta +
\sqrt{3} \sin \Delta ), \qquad y_u = 0,\eqno\new$$
where $\Delta$ is defined as
$$\Delta = \fr13 \acos \Bigl( { 3\over 2 \lambda} \Bigr)^{3/2}.\eqno\new$$
For convenience, let us define $x_\star = - x_u$. This quantity will
play an important r\^ole in what follows, so let us explicitly note that
it is always related to the scaled action of the unstable fixed point 
$\Jt_{\rm u}$ by  
\eqnam\ufpaction
$$\Jt_{\rm u} = \fr12 x_\star^2.\eqno\new$$
If we choose $D_1 <0$, then the separatrix looks like that of Fig.~7
with retrograde rotators below PS${}^\prime$ and prograde rotators
below PS. Then, by evaluating the actions of these two branches of the
separatrix, we find:
$$\eqalign{S_{\rm ret} &= \pi \lambda + 2\lambda \asin 
                      ( x_\star^{-3/2}) + {3(x_\star^3 -1)^{1/2} 
                      \over x_\star},\cr
S_{\rm prog} &= \pi \lambda - 2\lambda \asin ( x_\star^{-3/2} )
            - {3(x_\star^3 -1)^{1/2} \over x_\star},\cr
S_{\rm trap} &= 4\lambda  \asin ( x_\star^{-3/2} ) + {6(x_\star^3
-1)^{1/2} \over x_\star}.\cr}\eqno\new$$
Obviously, the area of the trapping region increases away from the tip.
The probability of capturing a retrograde rotator is
\eqnam\captureatu
$$P = {\displaystyle 2 \over  \displaystyle 1 + {\pi\over 2 \asin (
        x_\star^{-3/2} )}}.\eqno\new$$
This result was previously obtained by Yoder (1973) and Henrard
\& Lema\^itre (1983). Although the final expression is quite
complicated, the important point is that the capture probability
is a monotonically decreasing function of $\lambda$. As we move away
from the tip, the area of the trapping region increases but its
rate of swelling diminishes.

A problem which involves analysis at the U-point is that of
the two outer rings. Some barred galaxies like NGC 5701 or
A1340.6-2541 possess two outer rings of gas and stars (see
figures 2 and 5 of Buta 1986). Can stars be transferred between
the rings? As the pattern speed or the amplitude of the bar changes,
which one of the two rings grows? 

Fig.~7 shows an equiaction section at outer Lindblad resonance. The
innermost ring is built from stars moving on periodic orbits oriented
at right angles to the trough of the potential well (such as $\Ra$ on
Fig.~7). Likewise, the outermost ring corresponds to stars moving on
aligned periodic orbits $\Rb$ (see e.g., Athanassoula et al. 1982;
Athanassoula \& Bosma 1985). Suppose the pattern speed is diminished.
As the inertial response is negative, the position of exact resonance
moves outward (to higher angular momentum and therefore rightwards on
the equiaction section). This is indicated by the arrow on Fig.~7.
Then, the region of retrograde rotators (below ${\rm PS}^\prime$)
shrinks in size, while the region of prograde rotators (below ${\rm
SP}$) increases.  As a retrograde rotating star reaches the
separatrix, it may be captured into libration or escape into the
region of prograde rotation.  There is no transference of trapped
stars to prograde rotators, as the well is growing and the
adiabatically invariant action binds stars deeper with the separatrix.
In other words, all the transitions are from retrograde rotators to
trapped stars or to untrapped prograde rotators.  This is reversible,
so that if the pattern speed is increased, then all the transitions
are from trapped stars or prograde rotators to retrograde rotators.
This has the interesting consequence that stars cannot pass directly
from libration about $\Ra$ to libration about $\Rb$, and vice
versa. There is no direct exchange of stars from one ring to the
other. Stars from $\Ra$ pass straight across the trapping region and
end up untrapped, their eccentricities increased and rotating in the
opposite sense.

The bar can also grow or fade in strength. The effects of such changes 
are equivalent to changes in the pattern speed. This is because the 
capture probability (\captureatu) depends only on the parameter $\lambda$.
When $\lambda$ increases, the numbers of librators and prograde 
rotators increase at the expense of the retrograde rotators. When
$\lambda$ decreases, the situation is reversed. From the definition 
(\deflambda), we see that $\lambda$ increases if
\eqnam\lambdarule
$${-{\dot \Op}\over 2|D_1|(\Jr- \Jc)} - {\dot A_2 \over 3A_2} +{ {\dot
\Op} D_2\over 3 |D_1|^3}
> 0,\eqno\new$$
where $D_2 = {\partial^2 \Op / \partial \Js^2}$. The final term on
the left-hand side of (\lambdarule) is of higher order than the
remaining two and may be neglected. So, we deduce that increasing
the pattern speed $\Op$ or the bar strength $A_2$ both cause $\lambda$ 
to diminish. Notice this gives a seemingly paradoxical result. As the
bar grows, the numbers of trapped stars diminish! Normally, we expect 
growth of a potential well to enhance the probability of capture. This 
does not happen here because the U-point geometry tightly constrains
the possible change in the phase space area of the trapped stars.
So, if the bar is speeding up or increasing in strength, both the rings
fade. Conversely, if the bar is slowing down or dissolving, both the
rings grow.

There is one further deduction we can make. The capture probability
(\captureatu) is a monotonic decreasing function of $\lambda$. So, for
strong bars, the aligned outer ring is expected to be the most
prominent. For weak bars, the anti-aligned ring is the dominant
one. Although the sample is admittedly small, this appears to be borne
out by a visual examination of figure 5 of Buta (1986). In the more
weakly barred galaxies NGC 1291 and A0621.9-3211, the anti-aligned
ring is more conspicuous, whereas in the more strongly barred galaxy
A1056.3-4619, the aligned ring is the brightest. As the observational
evidence is suggestive rather than convincing, it would be interesting
to test this prediction with numerical simulations.

\subsection{3.3 Capture Probabilities at Non-Singular Points along
the Envelope}

\beginfigure{8}
\centerline{\psfig{figure=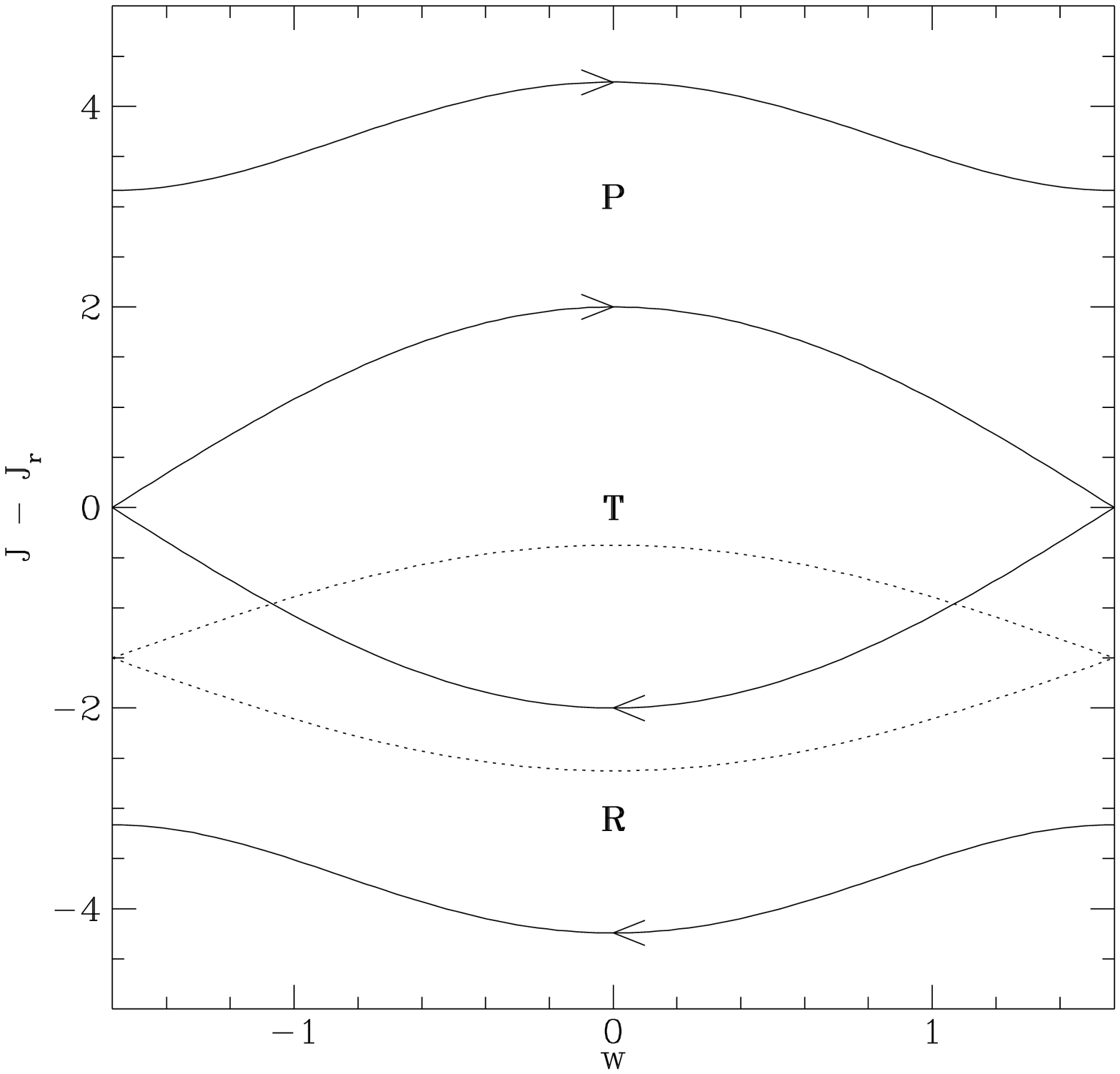,height=\hssize,width=\hssize}}

\caption{{\bf Figure 8.} A snapshot of phase space corresponding
to the Hamiltonian (3.27). The plot shows contours of constant energy
in the action-angle plane ($J,w$). The separatrix lobe (marked T) is 
moving down the diagram as well as changing in size. The region R 
(retrograde rotators) is shrinking, while the region P (prograde 
rotators) is expanding. The position of the separatrix lobe at a
later time -- when it has moved and changed shape -- is shown in the
dotted line.}
\endfigure

\noindent
In the previous examples, the scaling means that changes in pattern
speed and amplitude are coupled. At a general point on the envelope,
where the gradient term in (\oneharmonic) is less important,
we have to consider their variations independently. 

A rotating star can be captured into libration when the envelope 
of the section changes shape. Let us first consider the case when the 
trapping region is both moving and growing, but the lobe of the 
separatrix is symmetric. The Hamiltonian is
\eqnam\generalpoint
$$H = \fr12 D_1 {\hat J}^2 - A_2(1 + \alpha t) \cos 2w,\eqno\new$$
where ${\hat J} = J - J_r(t)$ is the action measured with respect to
the point of exact resonance. The amplitude $A_2$ is taken as positive
without loss of generality, but the inertial response $D_1$ may have 
either sign. The parameter $\alpha$ describes the steady expansion of 
the envelope. The resonant action is also assumed to be changing slowly 
and linearly with time like
$$\Jr (t) = \Jr(0) - \beta t.\eqno\new$$
This implies that the pattern speed is slowing down like
$${\dot \Op} = {\dot \Js}{\partial \Op \over \partial \Js} =
               -\beta D_1.\eqno\new$$
The equations of motion are
\eqnam\eqnsofm
$${\dot J} = - 2A_2 \sin 2w,\qquad {\dot w} = D_1(J - \Jr (0) + \beta
t).\eqno\new$$
For the moment, both $\alpha$ and $\beta$ are assumed small so 
adiabatic theory holds. 

The separatrix lobe (see Fig.~8) moves downwards and changes in
size. So, the regions P (corresponding to prograde rotators) and T
(trapped orbits) increase in size, whereas the region R (retrograde
rotators) diminishes. Initially, the separatrix lobe has an area equal
to $8 |A_2 / D_1|^{1/2}$. The change in this area is just
$${d S_{\rm trap}\over dt} = 4 \alpha \Bigl| {A_2\over D_1} \Bigr|^{1/2}.
\eqno\new$$
The change in the area P has two contributions. The first term
describes the increase due to the shifting potential well. The second
term is the diminuition caused by the encroachment of the separatrix
lobe T. This gives the formula
$${d S_{\rm prog} \over dt} = \pi \beta - \fr12 {d S_{\rm trap}\over dt}.
\eqno\new$$
Using (\capturehenrard), the capture probability is
\eqnam{\captureenvelope}
$$P =   {\displaystyle 2 \over  \displaystyle 1 + {\pi\over 2}{ \beta
         \over \alpha} \Bigl| {D_1 \over A_2} \Bigr|^{1/2}}.\eqno\new$$
If $\beta$ is negative, then the region T (see Fig.~8) moves upward.
Now, regions R and T increase in size, whereas P shrinks. The formula
(\captureenvelope) then gives the probability of capture of a prograde
rotator.

More realistically, we must deal with cases where the envelope
width and the inertia both depend on the action. In this case,
the Hamiltonian is 
$$H = \fr12 D_1 {\hat J}^2 + \fr16 D_2 {\hat J}^3 - 
(A_2 + B_2 {\hat J})( 1 + \alpha t) \cos 2w,\eqno\new$$
where ${\hat J} = J - J_r(t)$. Now, the separatrix lobe is asymmetric,
as the unstable fixed point is displaced from the stable fixed point
by $2A_2/D_1$ in action. The capture probability can be deduced by
perturbation methods when $D_2/D_1$ and $B_2/A_2$ are small.  We find:
\eqnam{\generaltw}
$$P = {\displaystyle 2 \over  \displaystyle 1 + {\pi\over 2} K},\eqno\new$$
where 
\eqnam{\capturedenominator}
$$K = {\beta\over \alpha}  \Bigl| {D_1 \over A_2} \Bigr|^{1/2}
      {1 \over {1 - {\beta\over \alpha}
      ({B_2\over A_2} - {D_2\over D_1})}} - {3B_2D_1 - D_2A_2\over
      6 |A_2D_1|^{3/2}}.\eqno\new$$

\beginfigure{9}
\centerline{\psfig{figure=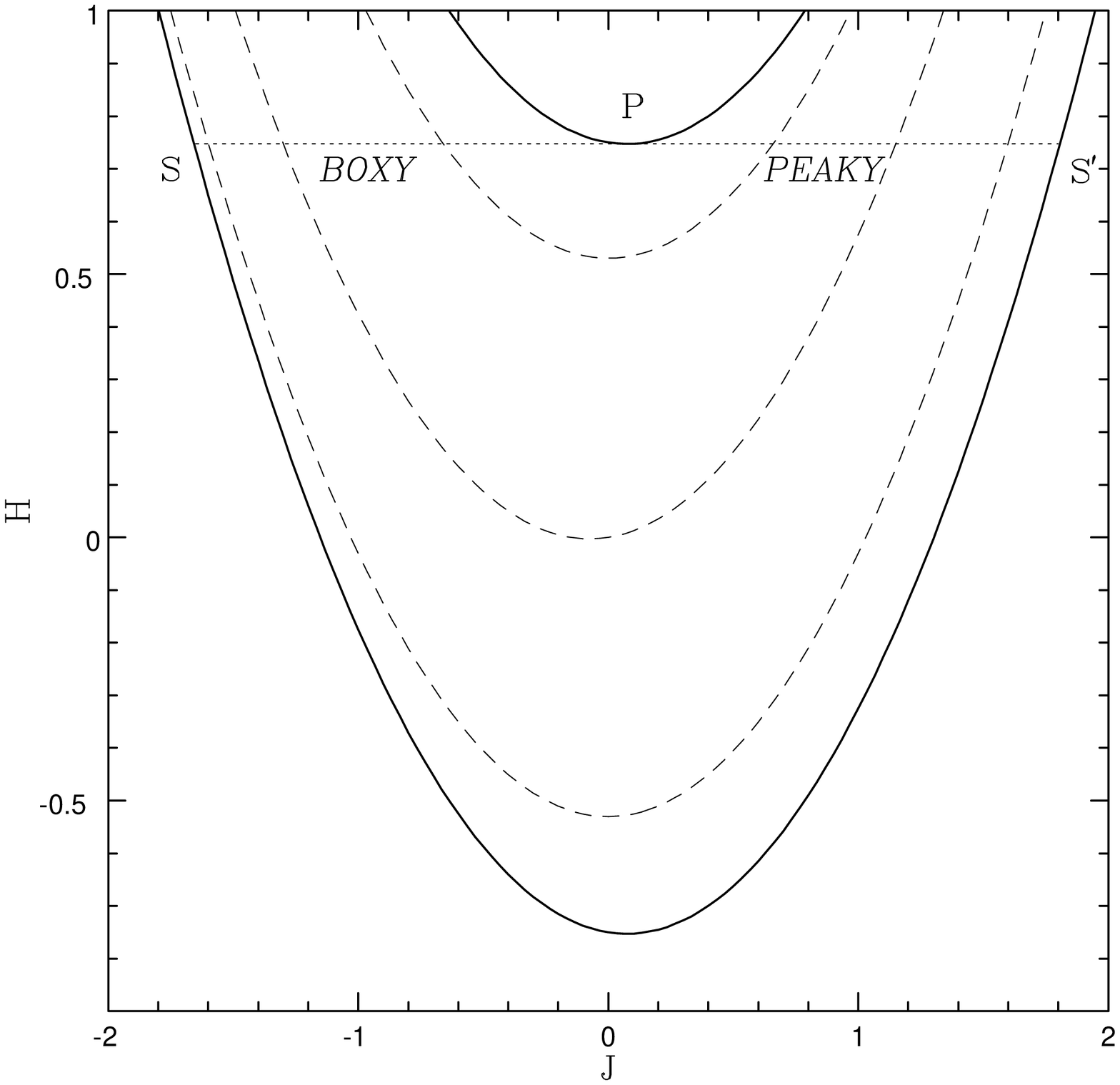,height=\hssize,width=\hssize}}

\caption{{\bf Figure 9.} An equiaction section corresponding to
the model Hamiltonian (3.39). The extra harmonic term leads
to a boxy well when ${\hat J} <0$ and a peaky well when ${\hat J} >
0$. The bunching of the inner lines show that the action of the
branch SP is reduced, whereas that of the branch PS${}^\prime$
is increased. This explains the enhanced capture of the boxy
well.}
\endfigure

\noindent
Tremaine \& Weinberg (1984) have already examined the case when the
potential well is moving but of fixed amplitude. Our formula
(\generaltw) reduces to theirs in the limit $\alpha \rightarrow 0$.
If $D_2$ is negative, the equiaction section becomes flatter as
we move towards increasing angular momentum. So, if the well is moving
towards high angular momentum, then the capture probability is
enhanced. We can see this in two ways. As the section becomes flatter,
the breadth in angular momentum of the trapping region has been
increased.  Equivalently, there is a smaller dispersion in resonant
frequency across the section so that outlying orbits are more readily
trapped. If $B_2/A_2$ is positive (and the well is moving towards
increasing angular momentum), capture is made more likely as we are
moving to a region of the section of greater width. In terms of the
asymmetry of the separatrix lobe, it helps to have the larger lobe in
the forward direction of the moving well.

The competition between terms in (\capturedenominator) again makes
clear that a growing instability need not be accompanied by a
monotonic growth in the membership of each resonant family. In other
words, although the growth of the well aids capture, this can be
offset by a movement of the resonance to a pattern speed (and hence
orbital shape) in which the orbit experiences a smaller averaged
potential.  When the pattern speed changes, the periodic orbit deduced
from (\eqnsofm) is actually offset from the bar potential by $\fr12
\asin( \beta /(2A_2))$. This is the means by which angular momentum 
is transferred between the periodic orbit and the well. This has the
following interesting consequence. Suppose a bar to be composed of
periodic orbits or boxes of small librational amplitude.  In a steady
state, the orbits are aligned and mutually provide the potential in
which they sit. When the bar is decelerated at a rate ${\dot \Op}$,
then we can anticipate a shear in the orientation of the orbits --
since $D_1A_2$ is different for each periodic orbit. Indeed, this can
even lead to escape when $D_1A_2$ is small. If the perturbation is
like a pure quadrupole, this condition may be best satisfied towards
the centre of galaxies or at the end of the bar.

\subsection{3.4 Refinements}

\beginfigure{10}
\centerline{\psfig{figure=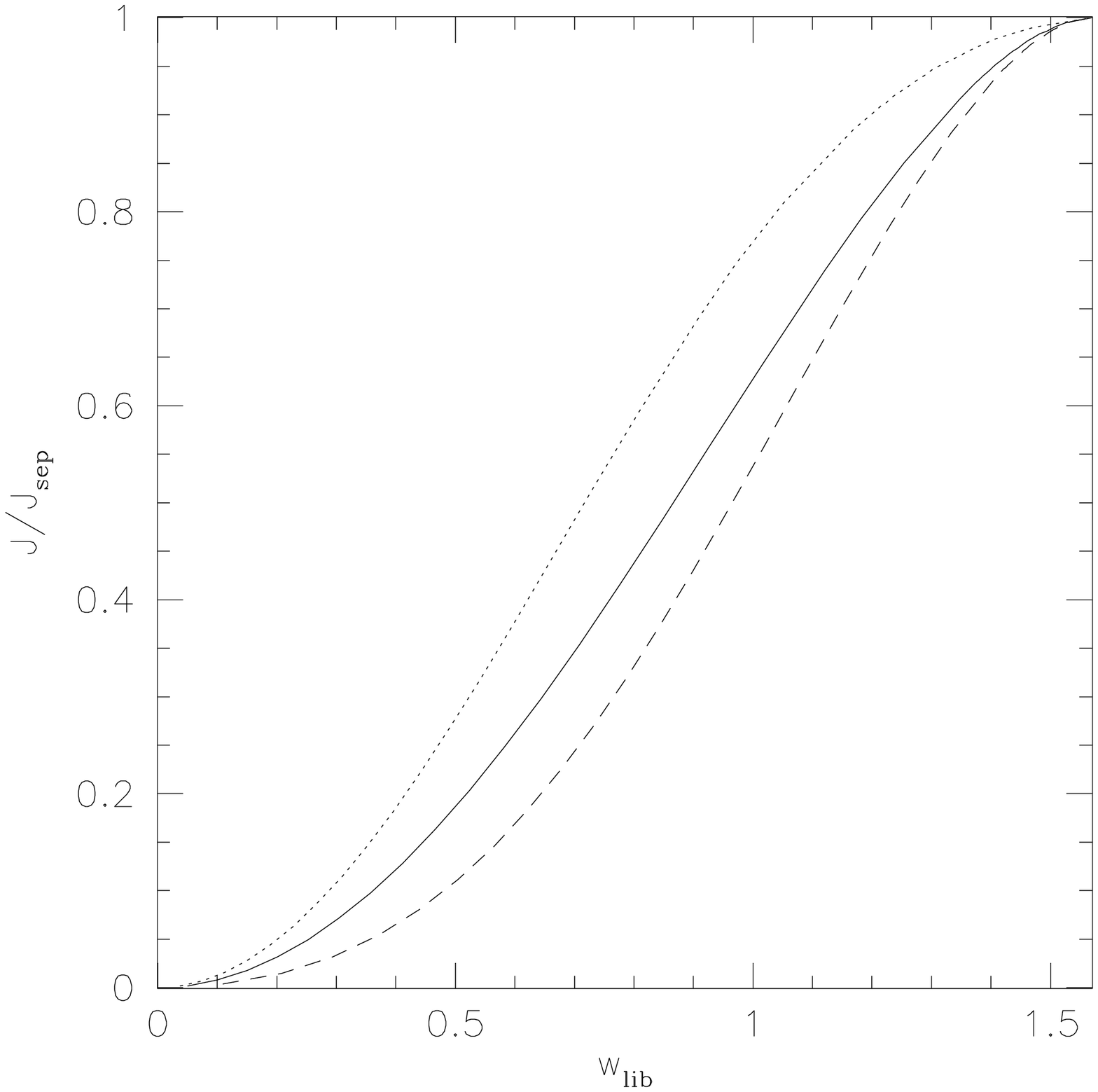,height=\hssize,width=\hssize}}

\caption{{\bf Figure 10.} After capture, the star falls
deeper into the well. The angular amplitude of libration
$w_{\rm lib}$ is plotted against the normalised action $J/J_{\rm sep}$.
The unbroken line is the $m=2$ harmonic well. The dashed line
is boxier ($A_4 = -0.20$) and the dotted line is peakier ($A_4 = 
0.20$). }
\endfigure

\noindent
The formulae presented in the previous sections employed
approximations to the Hamiltonian which we now relax. Here, we
investigate the effects of (1) the deviation of the section width from
the canonical form, (2) additional harmonics in the perturbation
potential and (3) spiral waves.

Examination of the exact sections at inner and outer Lindblad
resonance shows that the U-point approximation is valid in a region
very close to the circular orbit. As this tip constricts, we can use
the Hamiltonian:
$${\t H} = {\t J}^2 - 2 \lambda {\t J} - 2 (2 {\t J})^{1/2} 
(1 - \epsilon {\t J}) \cos {\t w},\eqno\new$$
which incorporates the contribution from the next order eccentricity 
term. The problem now depends on two parameters and does not possess
the simple scaling it had previously, which allowed us to consider 
pattern speed and growth changes as equivalent problems. What happens
to the capture probability as a steady resonance sweeps past? It
now becomes:
\eqnam\jimsheroics
$$P = P_0(x_\star) + \epsilon {\displaystyle \pi (8 - 5 x_\star^3)
        \over \displaystyle 4x_\star (x_\star^3 -1)^{1/2}
        ( \asin (x_\star^{-3/2})  + \pi /2 )^2 }.\eqno\new$$
Here, the first term $P_0 (x_\star)$ is the original capture
probability (\captureatu). A sketch of the derivation of the second
term is given in Appendix B, while $x_\star$ is defined in terms of
the action of the unstable fixed point by (\ufpaction).  This is the
same as used in Section 3.2, but of course the position of the
unstable fixed point itself has moved in the perturbed problem.  Note
that $x_\star >1 + 2\epsilon$, so that the singularity in the second
term in (\jimsheroics) never occurs in physical applications where the
tip constricts. The probability of capture close to a U-point can be
either increased (if $x_\star < 2/5^{1/3}$) or decreased (if $x_\star
> 2/5^{1/3}$), assuming $\epsilon$ is positive. When $x_\star$ is
large, the width of the envelope of the section becomes more uniform
and the phase area of the trapped orbits changes more slowly. When
$x_\star$ is small, the unstable fixed point lies close to the
tip. The perturbing term has a larger relative influence on shrinking
the area near the tip than the area of trapped orbits -- and so the
capture probability is increased.

More serious, perhaps, is the neglect of higher harmonic terms.  This
was already clear in the analysis of the Pluto-Neptune system in
Section 2.3. As the perturbation grows in size or the orbit becomes
more eccentric, the influence of the higher harmonics becomes more
important. This can manifest itself in two ways -- first, by changing
the capture probability, and second, by modifying the response density
contributed by the captured star. The effect of the harmonics is
different at the three r\'egimes of the equiaction section. The
simplest r\'egime to consider is at a general point on the section,
where the Hamiltonian may be approximated by
$$H = \fr12 D_1 {\hat J}^2 - \Bigl[ A_2 \cos 2w + A_4 {\hat J} 
\cos 4w \Bigr],\eqno\new$$
where $A_4$ is assumed small and positive. The corresponding section 
is illustrated in Fig.~9. The capture probability -- which would 
vanish in the absence of the higher harmonic term -- is now
$$P = {16\over 3\pi}{A_4 \over |A_2 D_1|^{1/2}},\qquad \qquad A_4 D_1 >0,
\eqno\new$$
when the well is moved towards the boxy part of the section (${\hat J}
<0$). The probability remains zero when the well is moved in the
opposite direction -- indeed, now trapped stars seep out of the
well. This may be understood on recalling the result of Section 3.3,
namely that capture is enhanced by having the larger lobe in the
forward direction of the moving well. Armed with this result, 
we may be tempted to believe that boxy wells always aid capture. This 
is not the case, as we shall now show. At the U-point, the second 
harmonic gives a contribution that goes like the square of the 
eccentricity. So, the U-point Hamiltonian is modified to:
$${\t H} = {\t J}^2 - 2 \lambda {\t J} - 2 (2 {\t J})^{1/2} 
\cos {\t w} - \mu {\t J} \cos 2{\t w},\eqno\new$$
which effectively introduces an octopole into the bisymmetric wave.
The calculation is outlined in Appendix B and the result for the
capture probability is:
\eqnam\jimsdeeds
$$P = P_0(x_\star) + \mu{\pi x_\star (x_\star^3 -1)^{1/2} \over
6 (\asin (x_\star^{-3/2})  + \pi /2)^2} .\eqno\new$$
Surprisingly, when the well is sharpened ($\mu >0$), the probability 
of capture is enhanced.

The effects of the extra harmonic term on capture can therefore
be quite subtle. Let us now turn to our second point of how quickly
a captured star sinks into the well. We take the Hamiltonian to be 
a modified form of (\generalpoint)
$$H = \fr12 D_1 {\hat J}^2 - \Bigl[ A_2 \cos 2w + A_4 \cos 4w 
\Bigr].\eqno\new$$
After capture, the star descends deeper into the well. The action of
its new libration is again adiabatically invariant. This can be
exploited to give a rough way of assessing when its response density
becomes supportive of the trapping potential. In Fig.~10, the angular
amplitude of libration $w_{\rm lib}$ is plotted against the action of
the orbit normalised to the action of the separatrix $J/J_{\rm
sep}$. The unbroken line refers to a pure $m =2$ harmonic well, the
dashed and dotted lines to boxier and peakier wells respectively. The
peakier well ingests the star more quickly. The figure makes clear the
relative change in action needed for the star to reinforce the
perturbation well.

\beginfigure{11}
\centerline{\psfig{figure=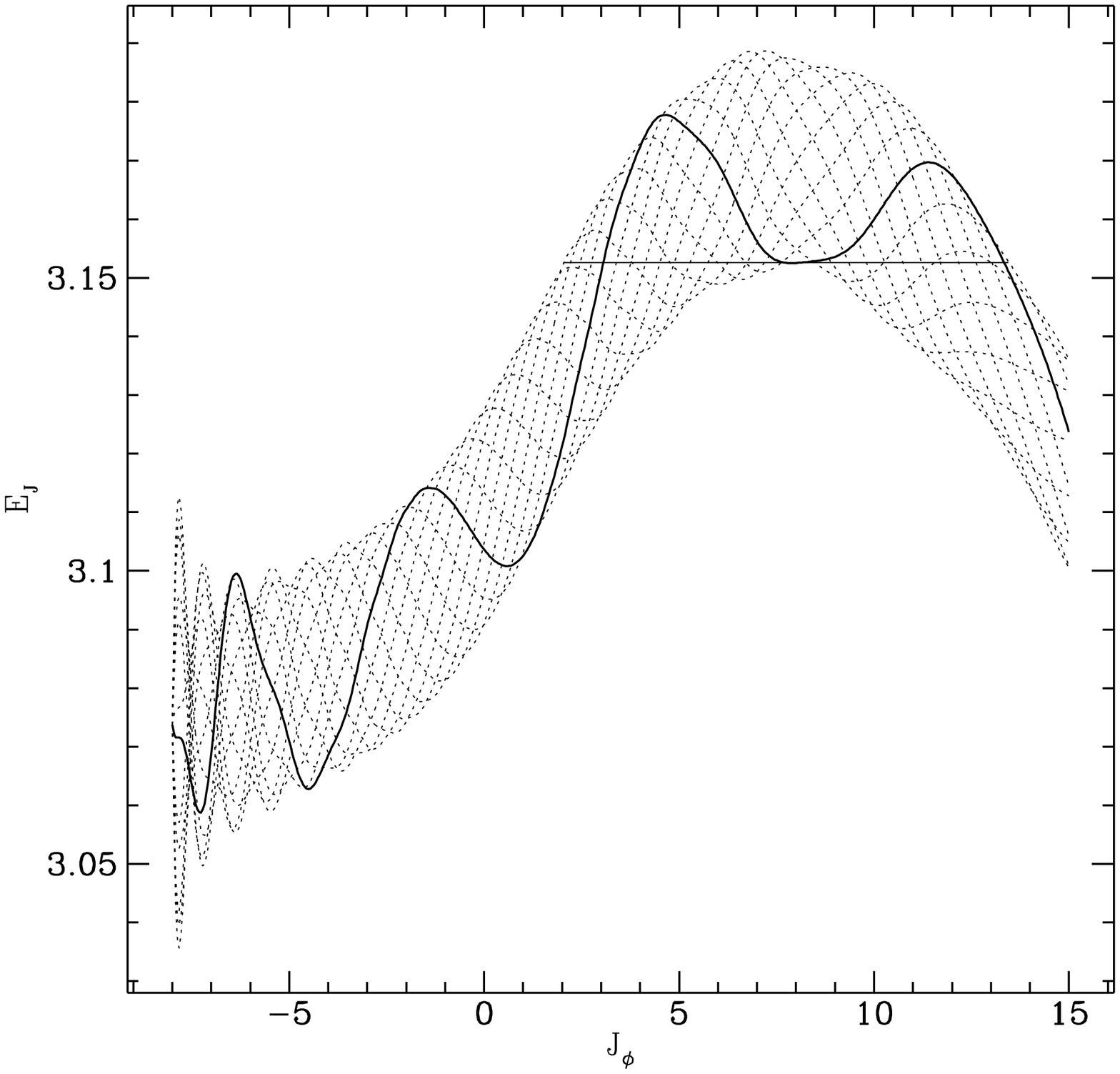,height=\hssize,width=\hssize}}

\caption{{\bf Figure 11.} An equiaction section for a tightly
wound spiral wave. The lines representing the values of the Jacobi
integral at constant azimuth are shown in dotted lines -- one is
highlighted for clarity. The solid horizontal line is the separatrix.
Interestingly, the separatrix now corresponds to a libration with
amplitude {\it greater} than $\pi /2$. [The axisymmetric model used is
the cored Mestel disk of Fig.~3. It is subjected to a logarithmic spiral 
perturbation with $m=2$ and radial wave number $\alpha \sim 17$. This 
corresponds to a pitch angle of $\sim 83^\circ$, so that the spiral 
is tight].}
\endfigure

\beginfigure{12}
\centerline{\psfig{figure=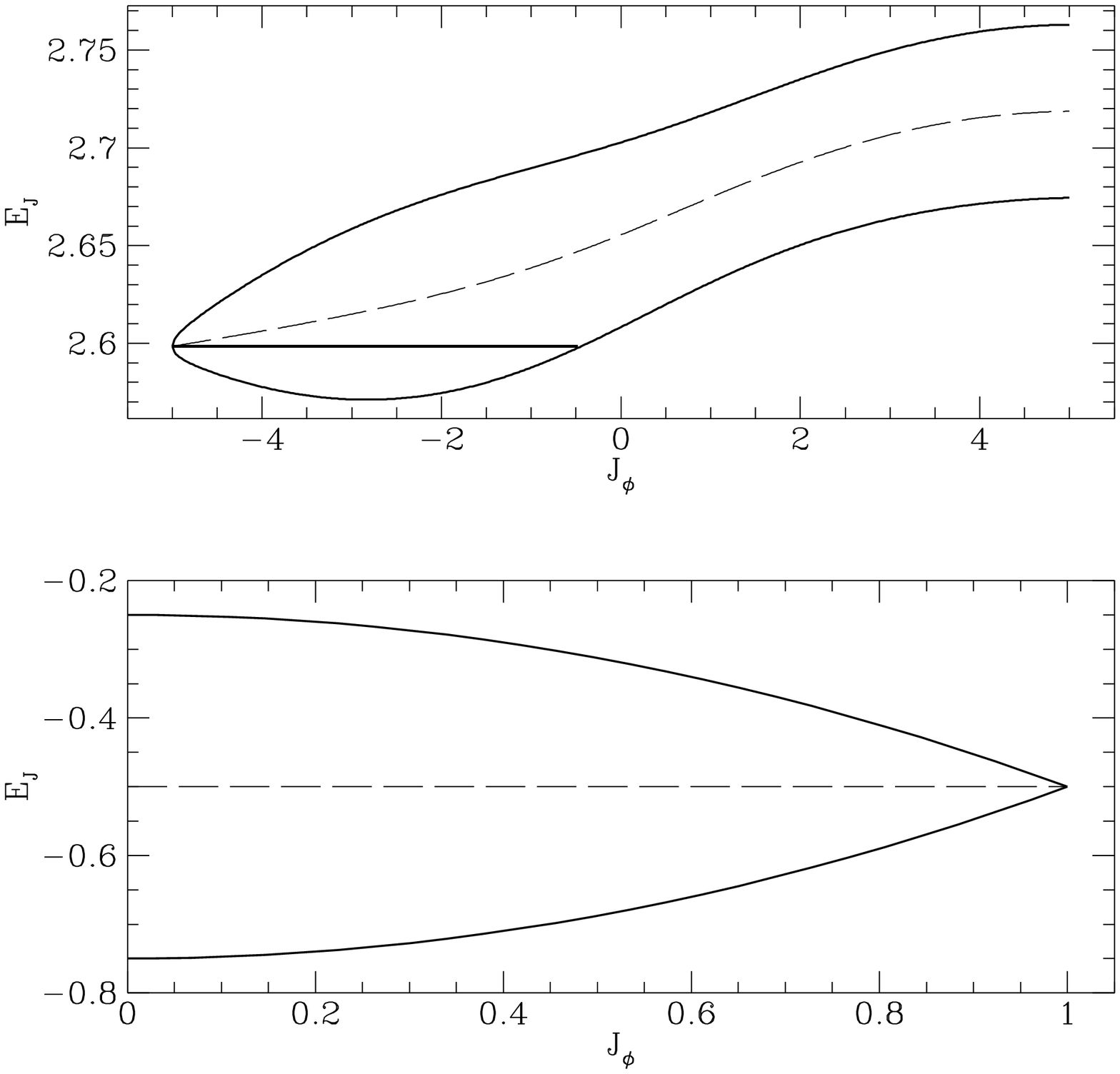,height=\hssize,width=\hssize}}

\caption{{\bf Figure 12.} The upper equiaction section shows the
negative angular momentum branch for the corotation resonance. The bold
horizontal line corresponds to a librating orbit that almost reaches
to the galactic centre. The overall flatness of the section means
that small disturbances can lead to large changes in orbital shape.
The lower panel is an equiaction section for Keplerian ellipses
subjected to a stationary planar quadrupole. The section is remarkable
for its flatness. The central seam is horizontal manifesting the
degeneracy of the Keplerian orbits.}
\endfigure

\beginfigure{13}
\centerline{\psfig{figure=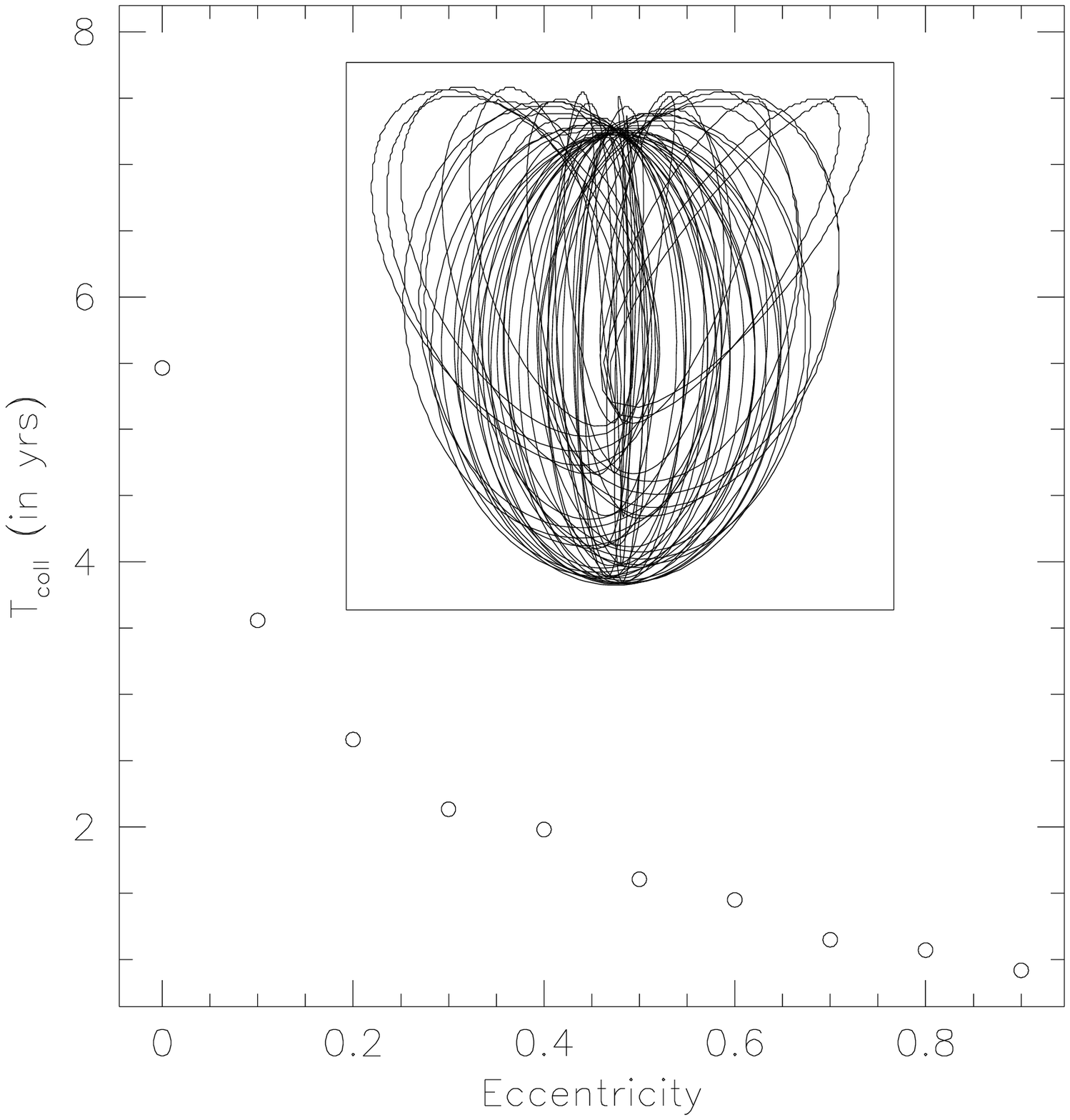,height=7.5truecm,width=\hssize}}
\vskip\hssize

\caption{{\bf Figure 13.} The two panels show how the collision
time $T_{\rm coll}$ of a lunar orbit varies with eccentricity and 
orbital inclination. The inset to the first panel shows the lunar orbit 
in the Lidov experiment. The inset to the second panel depicts the 
orbital geometry.}
\endfigure

Spiral waves change the section in two ways. First, the lines of
constant slow angle now oscillate within the envelope. Second, the
spiral wave effectively shears the potential in annuli so that the
average potential has to be re-computed for each periodic orbit. It is
not hard to see that the first change within an envelope of fixed
shape leaves the capture probability formulae unaltered.  The only
modifications we need to consider are due to the second.  The barred
potential of Fig.~3 is replaced by a tightly wound logarithmic
spiral in Fig.~11, viz;
$$\psi_{\rm p} = \epsilon {R^2\over {\t R}_c^2 + R^2} \cos ( m\phi 
-\alpha \log r).\eqno\new$$
The spiral wave has removed the X-points and -- as we shall see in 
Section 4.2 -- this permits angular momentum transport throughout 
the section. The Hamiltonian close to exact resonance may be written as
$$H = \fr12 D_1 {\hat J}^2 - A_m \cos m (w -  \lambda {\hat J}),\eqno\new$$
where $\lambda$ is a measure of the spirality and strictly speaking
depends on the fast action $\Jf$. In Fig.~11, the highlighted line has
a positive curvature near the position of exact resonance, whilst the
envelope of the section has negative curvature. This is characteristic
of strong spirality and happens whenever $m |\lambda^2 A_m / D_1
|^{1/2} > 1$. There is no distinction between trailing and leading
spiral waves as far as capture and escape are concerned.

\eqnumber =1
\def\chaphead{\hbox{4.}}
\section{4 Applications}

\subsection{4.1 Large Eccentricity Changes}

The equiaction section allows us to find the periodic orbits on which
cold gas settles. As the periodic orbit increases in eccentricity, it
develops self-intersections and can no longer support a steady gas
stream. However, this is not true of one orbital family, namely the
($\ell = 0, m= 1$) family. All the orbits belonging to this sequence
are not self-intersecting. In principle, gas could be shipped right to
the galactic centre through a continuous series of periodic
orbits. This provides one possible mechanism for fuelling the nucleus
with counter-rotating gas from large radii. Of course, streams of
counter-rotating gas are common in spirals and S0s (e.g., Bertola,
Buson \& Zeilinger 1992). This mechanism is viable if the angular
momentum of the periodic orbit changes markedly under moderate
disturbances. This is true of any resonance possessing a flat section.
An example of this is obvious from Fig.~1, where the ($\ell = 0,
m=1$) fast action tangents are nearly parallel to the resonance
lines. This situation persists over a broad range of pattern
speeds. The $m=1$ disturbance couples most strongly to this set of
resonant orbits and the U-point at its tip ensures that there are
always nearly circular periodic orbits. In an $m=2$ disturbance, the
tip becomes a V-point, generally without an associated stable periodic
orbit for the gas to sit on. The only stable periodic orbit is that
associated with the moving resonance and this is capable of moving
from circular to radial orbits. Stars, on the other hand, can undergo
the large scale shape changes through libration alone, a typical
example of which is indicated on Fig.~12. This shows the negative
angular momentum branch for the co-rotation ($\ell =0, m=1$) resonance.
The axisymmetric model is the cored Mestel disc subject to the 
disturbance
$$\psi_{\rm p} = \epsilon {R\over R^3 + a^3}\cos m\phi.\eqno\new$$ 
As indicated by the bold horizontal line on the upper panel of Fig.~12,
the libration swings from circular to very nearly radial for some
orbits.

Close to the galactic centre, if a massive black hole is present, the
gravity field is nearly Keplerian. The gas can follow the closed
periodic elliptical orbits. In order to direct gas and stars onto 
the hole, a mechanism of increasing the eccentricity of the orbit is again
required. A tidal field inclined at an angle to the orbital plane is
one possibility. We can estimate the timescale on which the mechanism
proceeds as follows. Let us consider the Hamiltonian
\eqnam\crudeham
$$H = \fr12 D_2 {\hat J}^2 - A_2({\hat J},t) \cos 2w.\eqno\new$$
For the Kepler potential, $D_2$ vanishes for the ($\ell =0, m =1$) 
resonance. The amplitude can be taken as (c.f., the lower panel of
Fig.~12)
$$A( {\hat J},t) = a(t) \Bigl[1 - { {\hat J}^2\over 
{\hat J}_{\rm circ}^2} \Bigr],\eqno\new$$
where $a(t)$ is an arbitrary function of time. Let us
choose the normalisation so that the circular orbit has unit action,
i.e., ${\hat J}_{\rm circ} = 1$. The trajectories of the equations of
motion are independent of the function $a(t)$, namely
$$\cos 2w \propto { 1\over 1 - {\hat J}^2}.\eqno\new$$
A particularly simple case is when $a(t)$ is just a constant, say $A$.
Then, gas starting on an orbit with action-angle coordinates 
$({\hat J}_0, w_0)$ loses all its angular momentum in the time
\eqnam\crudetime
$$T_{\rm coll} = {1\over 2A} \int_0^{{\hat J}_0} {d\,{\hat J} \over
( [1-{\hat J}^2]^2 - [1-{\hat J}_0^2]^2 \cos^2 2w_0 )^{1/2}}.\eqno\new$$
This expression can be recast as an elliptic integral, but it is easy
to work out numerically as needed. If the gas starts out on exactly
circular orbit, then it can never lose all its angular momentum, but
small eccentricities can be amplified quickly. In his fascinating
book {\it Huygens \& Barrow, Newton \& Hooke}, Arnold (1990) briefly
reported an observation made by Lidov (1963).  Lidov discovered that
if the orbit of the Moon is turned through $90^\circ$, its
eccentricity increases so rapidly under the action of the tidal forces
of the Sun that it collides with the Earth in four years! An
order-of-magnitude confirmation is provided by (\crudetime) with the
constant $A$ roughly equal to the tidal potential due to to the Sun,
i.e., $A \sim G M_\odot r^2 / R^3$. Here, $R$ of course refers to the
Earth-Sun distance and $r$ to the Earth-Moon distance. Taking the
eccentricity of the Moon's orbit at the present day as $0.055$ (Allen
1973), then the time taken for collision with the Earth $T_{\rm coll}$
is calculated by (\crudetime) as $\sim 3.7$ years.

To investigate this problem further, the equations of motion of the Moon 
in axes rotating with the Earth and under the action of the solar tides
$$\eqalign{ {\ddot x} =& -{GM_\oplus x\over r^3} + {3GM_\odot x\over R^3} 
+ 2 \Omega {\dot y}, \cr
{\ddot y} =& -{GM_\oplus y\over r^3} - 2\Omega {\dot x},\cr
{\ddot z} =& -{GM_\oplus z\over r^3} - {GM_\odot z\over R^3}.\cr}
\eqno\new$$
were integrated numerically by fourth-order Runge-Kutta methods. Here,
the $x$-axis points towards the Sun, and the $z$-axis points out of
the plane of the ecliptic. Fig.~13 shows the results of these
integrations, in which the Moon always starts off on the z-axis at
apocentre. Coriolis force causes the orbital plane to precess. The
conserved perturbation potential that is appropriate here is an
azimuthal mean of the tidal potential felt around the figure of the
orbit. In the Lidov experiment (shown as an inset), this means that
the line of apsides of the orbit eventually settles to a torquing
angle $\theta_{\rm t}$ of $\sim 51^\circ$ (in contrast to the simple
Hamiltonian (\crudeham), where $\theta_{\rm t} = 45^\circ$). The final
torquing angle for arbitrary eccentricity may be deduced from the
conserved potential as
$$\cos 2 \theta_{\rm t} = \fr15 (6e^2 -1).\eqno\new$$
The maximal torque occurs when $\theta_{\rm t} = 45^\circ$ so that the
eccentricity curve flattens out. When the orbital plane is inclined,
the collision time $T_{\rm coll}$ increases dramatically, and this
limits the range of orbits that can be tidally elongated. This
mechanism may have interesting applications in the central regions of
galaxies dominated by black holes, where the potential is nearly
Keplerian. Tidal forces, perhaps caused by a sinking object, can drive
orbiting stars into the hole. Another application is to the survival
of high inclination comets and asteroids, where indeed this
instability has already been discovered anew (e.g., Kozai 1980; Stagg
\& Bailey 1989).

\beginfigure{14}
\centerline{\psfig{figure=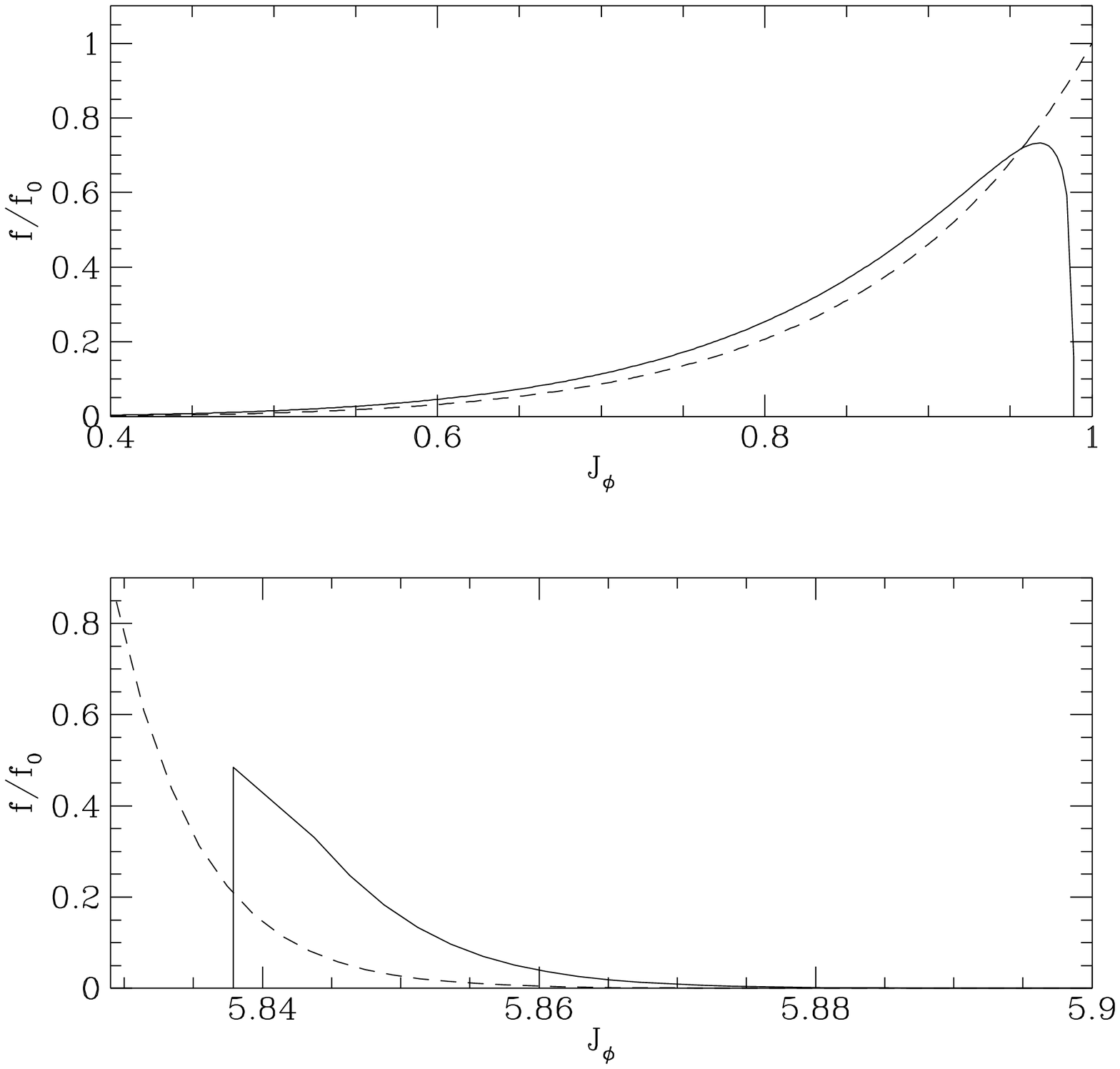,height=\hssize,width=\hssize}}

\caption{{\bf Figure 14.} The upper and lower panels show the distribution 
function of the warm exponential disk at inner and outer Lindblad resonance 
respectively. In each case, the distribution function is shown
before (in dashed line) and after (in unbroken line) a wave sweeps across 
the equiaction section. The disturbance heats the disk and opens up a 
hole in the distribution function corresponding to the missing circular 
orbits. [The upper panel is drawn with $\Jf = 0.5$, the lower panel
with $\Jf = -2.9$ in units in which $\sigma_0 = 1/4$ and $v_0 = \alpha =
1$.] }
\endfigure

\beginfigure{15}
\vskip\hssize

\caption{{\bf Figure 15.} The creation of a notch in the distribution
function. The upper panel highlights two strips of phase. The dark phase
is denser. After passage across the separatrix, stars from these strips
are partitioned between two populations as shown in the second panel.
The region of lower phase density now lies between two high density
strips. (Here, the phase density is stratified on lines of
constant $E_{\rm J}$. The equiaction section does not justly represent
phase areas, but for pictorial convenience, we have drawn each strip
with a constant strip-averaged phase density).
}
\endfigure

\subsection{4.2 Modes}

A disturbance changes the angular momentum of the stars in two ways.
First, it may capture a star and carry that trapped star as it moves.
Otherwise, it will {\it flip} the stars that it encounters from one
sense of rotation to the other. The angular momentum transferred in
the second process depends upon the speed of transition. Tremaine \&
Weinberg (1984) investigated this for non-singular points along the
envelope. Here, we supplement their calculations with those
appropriate to orbits of low eccentricity close to a U-point and
general orbits near X-points.

Now let us imagine a disturbance at inner Lindblad resonance, the
pattern speed of which is slowly decreasing. As the disturbance sweeps
across the section, it encounters first the low phase density tail of
this distribution. If we assume the amplitude of the wave to be
steady, then each star is flipped and none are caught (see Section
3.2). This angular momentum jump decreases as we approach the tip of
the section, but all stars are shifted and so, after the disturbance
has passed, a hole has opened up with a sharp edge and the
distribution has lost the orbits that were originally very near to
circular. In addition, the tail of the distribution has been
distended. For example, suppose we have a warm exponential disk with a
radial scale-length $1/ \alpha$ and a central velocity dispersion
$\sigma_0$. If the disk has a completely flat rotation curve ($v_{\rm
circ} =1$), then the potential in the plane is that of Mestel
(1963). If the velocity dispersion is also exponentially declining,
then the distribution function has the form (Newton 1986; Binney 1987;
Kuijken \& Tremaine 1991)
$$f(J_R, J_\phi)  = f_0 \exp \Bigl[- \sigma_0^{-2}\exp( \alpha J_\phi)
 \kappa J_R\Bigr], \eqno\new$$
where $f_0$ is independent of the velocities. Along the section, this
becomes
$$f(\Jf,\Js) = f_0 \exp\Bigl[ -\sqrt{2} \sigma_0^{-2} \exp( \alpha \Js ) 
({\Jf\over \Js} \pm {1\over 2} ) \Bigr].\eqno\new$$
Here, the negative sign is appropriate for ILR, the positive sign for OLR.
For a flat rotation curve, the Hamiltonian for epicyclic orbits
may be approximated by
$$H = {1\over 2} + \log J_\phi + \sqrt{2}{J_R\over J_\phi} - {11J_R^2\over
12 J_\phi^2}.\eqno\new$$
from which the inertial response $D_1$ at both ILR and OLR is
negative. This can also be deduced by inspection of the lower panel of
Fig.~1. The distribution function is so steep that we assume for
simplicity that $D_1$ is the same for all the stars within each
section and the Hamiltonian can (with a scaling) be brought into the
form (\upointham). To work out the flip experienced by each star, we
match its initial action to the action along the incident branch of
the separatrix. The corresponding jump is then just the phase area of
the separatrix lobe. As before, let the scaled action of the unstable
fixed point P at the moment of flipping be $x_\star^2/2$.  The phase
area between any orbit and the circular orbit at the end of the tip is
$\pi |\Jc - J|$. This is scaled and matched to the area under the
inner separatrix branch
$$S_{\rm ret} = ({\pi\over 2} - \asin (x_\star^{-3/2}) )
( x_\star^2 + {2\over x_\star}) - {3(x_\star^3 -1)^{1/2} 
\over x_\star}.\eqno\new$$
The flip $F$ will then be 
$$F = 2\Bigl[ ( x_\star^2 + {2\over x_\star}) \asin(x_\star^{-3/2}) +
{3 (x_\star^3 -1)^{1/2} \over x_\star}\Bigr].\eqno\new$$
The final area under the phase curve $S_{\rm prog}$ is just the initial
area plus the flip, or
$$S_{\rm prog} = F + S_{\rm ret} = \pi \Bigl[ x_\star^2 + {2\over x_\star}
\Bigr] - S_{\rm ret}.\eqno\new$$
When the disturbance has moved far away, this is -- once unscaled -- the 
new slow action or angular momentum $J_\phi^{\rm n}$ of the star. The final
coarse--grained distribution function along the section is
$$f_{\rm coarse}(J_R,J_\phi^{\rm n}) = f(J_R,J_\phi){{\rm d} J_\phi \over 
{\rm d} J_\phi^{\rm n}},  \eqno\new$$
where the Jacobian is 
$${{\rm d} J_\phi \over {\rm d} J_\phi^{\rm n}} = { \fr{\pi}2 -  \asin
                      ( x_\star^{-3/2}) \over \fr{\pi}2 +  \asin
                      ( x_\star^{-3/2})}.\eqno\new$$
The distribution function is coarse--grained because \lq\lq air" has
become mixed up with the phase as the well passes and the phase area
deflates. This point is made clearly by Sridhar \& Touma (1996), who
performed a related calculation on vertical heating of stars in a
galactic disk by sweeping resonances. Fig.~14 shows the initial
(dashed line) and final (unbroken line) distribution functions for the
warm exponential disk. The upper panel refers to ILR, the lower panel
to OLR. In each case, the flips get smaller towards the tip and the
last flip creates a hole with width $W$
$$W = {3\pi\over 2^{1/3}} \Bigl| {A_2\over D_1} \Bigr|^{2/3}.
\eqno\new$$
The hole has been advected by the well from a region of zero phase 
density and deposited at the end of the tip entirely vacating the
near-circular orbits. If the disturbance moves in the opposite direction,
{\it into} the section, a different redistribution of phase occurs.
The section bends as shown in Fig.~15 and the phase settles into
horizontal layers until the appearance of the separatrix. Layers
above the separatrix at this point will then be processed in turn.
A portion of the phase will move into the main well and a portion
will flip in accordance with the capture probability. As the diagram
makes clear, a notch is created in the distribution function. Adiabatic
transfer of phase is one mechanism for creating inverted populations --
in the sense that the phase space density gradients have their sign
changed. This is true here but the phase density is still most 
concentrated about the two periodic orbits. Such holes and notches 
in the distribution function are well-known sites of instability 
(Toomre 1981; Sellwood \& Kahn 1991).  

\beginfigure{16}
\centerline{\psfig{figure=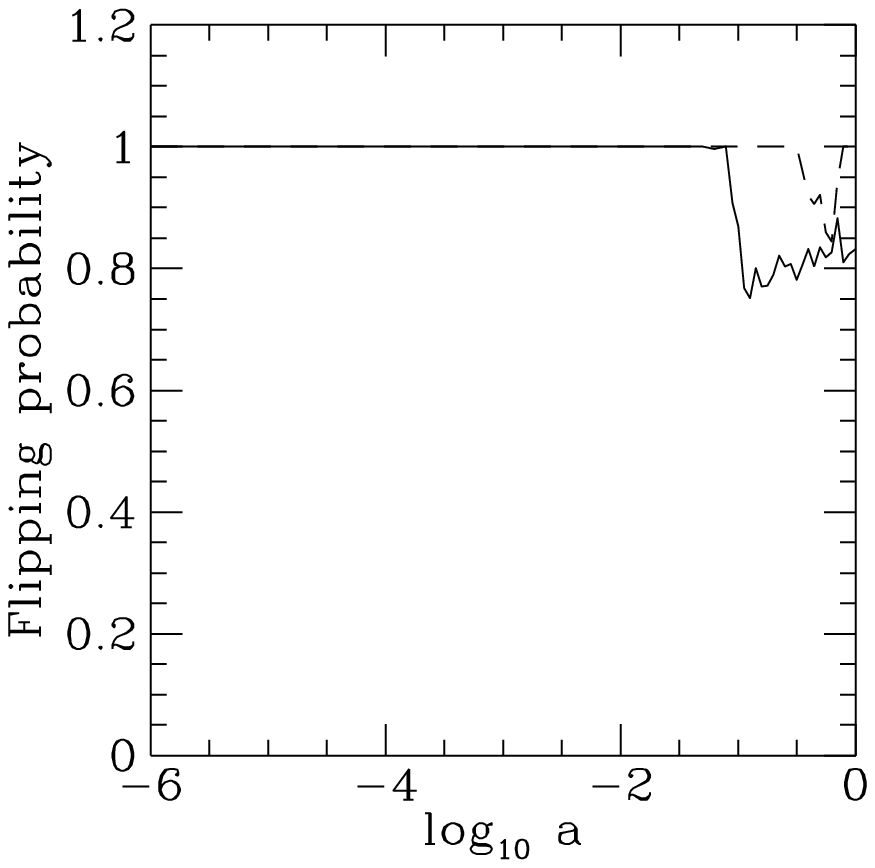,height=6truecm,width=6truecm}}

\caption{{\bf Figure 16.} Monte Carlo simulations of the equations
of motion at a U-point (3.16) subject to a slow turn-on of the trapping
potential. In adiabatic theory, all the stars flip from prograde to 
retrograde rotation. The probability of flipping is plotted against 
the constant $a$. The larger the value of $a$, the faster the speed 
of separatrix crossing. Adiabatic theory is seen to hold good over 
nearly five orders of magnitude. (The full line refers to $\lambda_0
= 3$, the dashed line to $\lambda_0 = 16.5$).}
\endfigure

The holes lead to a systematic gain in angular momentum at OLR,
and a loss at ILR. This is similar to, but not identical with, the
effect discovered by Lynden-Bell \& Kalnajs (1972). Their famous formula 
for angular momentum transport applies to infinitesimally small waves,
so small in fact that no star completes a traverse of the
separatrix. Consequently, the issue of capture and escape is
irrelevant to their calculation as is suggested by the absence of
terms involving the gradient of the amplitude of the disturbance,
i.e.,
\eqnam\lbk
$${\dot J_\phi} = {-1\over 8\pi} \int d\Jf \int d\Js m^2 {\partial F \over 
\partial \Js} A_m^2 \delta( m\Op - m\Omega - \ell \kappa).\eqno\new$$
What is relevant, however, are the amounts of phase just entering into 
libration near the two branches of the separatrix and this is manifest 
in the gradient of the distribution function in (\lbk). In the flipping
mechanism, the hole displaces phase around itself, irrespective of density
gradients. A nice picture to have in mind is of a bubble in a spirit
level.

The shunting mechanism relies on there being no capture and so the
hole remains empty. It is interesting to see if this situation
persists when the amplitude of the perturbation is turned on more
quickly. To this end, Monte Carlo simulations were performed with the
equations of motion derived from the Hamiltonian (\upointham) with a
Gaussian turn-off
$$\lambda = \lambda_0 \exp (-a t^2),\eqno\new$$
From (\deflambda), turning $\lambda$ off is equivalent to increasing
the amplitude of the disturbance.  The constant $a$ determines the
speed of separatrix crossing.  The results are shown in Fig.~16 for
two different values of $\lambda_0$. The probability of flipping
remains unity as we move out of the adiabatic r\'egime. The dip has an
interesting explanation. The turn-off law has an inflection point at
$t = (2a)^{-1/2}$. If stars make their inward crossing of the
separatrix just before this time, they can then lose more energy than
they gain on the outward crossing. This is a deviation from adiabatic
theory. Whilst subtleties of this kind should be observed, adiabatic
invariants are -- in the nice phrase of Alar Toomre -- \lq\lq very
forgiving".

\beginfigure{17}
\centerline{\psfig{figure=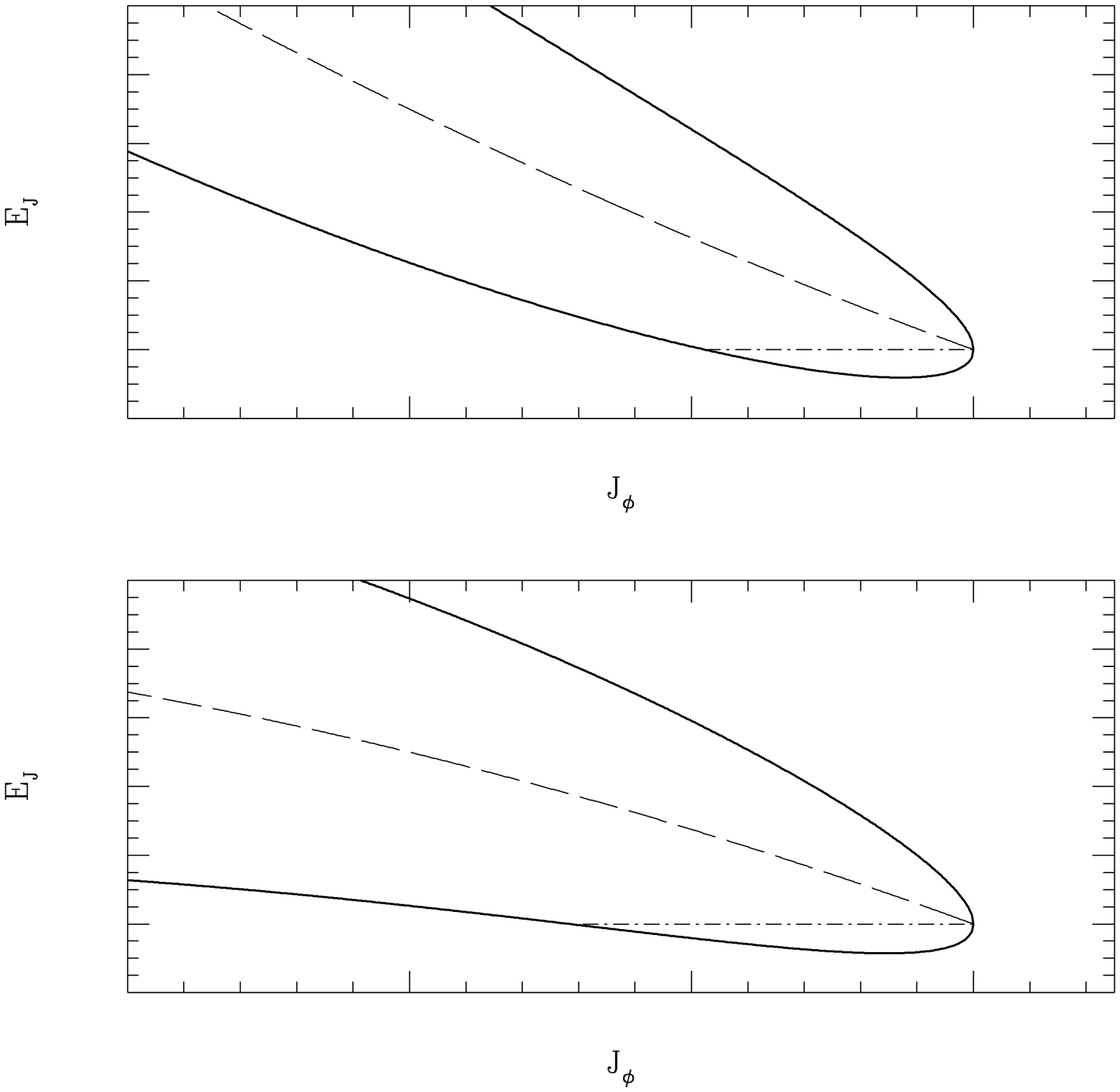,height=\hssize,width=\hssize}}

\caption{{\bf Figure 17.} Equiaction sections for stars whose ILR
frequencies are below that of the applied perturbation. The distorted
circular orbits are trapped into libration below the pseudo-separatrix
indicated by a dot-dashed line. In the upper panel, the stars possess
positive inertial response, in the lower panel, negative. Capture into a
potential reinforcing libration is now assisted by negative inertial
response, the curvature of the section acting with the swelling of the
envelope}
\endfigure

\noindent
Since capture at a U-point even by a rapidly growing wave is hard, it
would appear difficult to establish a density supporting population to
build, for instance, a bar from near-circular ILR orbits. This
difficulty can be circumvented by capture at frequencies above the
ILR. In this case, the trapped family are the distorted circular
orbits and, intriguingly, a negative inertial response now promotes a
quick shedding of angular momentum and a density enhancing elongation
of the orbits. This is illustrated on the equiaction sections shown in
Fig.~17, where the curvature of the section in the negative inertial
case acts with the swelling of the envelope.

The closing of the envelope at the U-point led to a hole in the
distribution function. Similar behaviour occurs close to an X-point.
If we consider the transition between the two main panels of Fig.~6,
stars close to the X-point and above S$\Pa$ are again dispatched into
counter-rotation, with their angular momentum flip diminishing the
closer they are to the X-point. Eventually, the separatrix is replaced
by a pseudo-separatrix. The advection of angular momentum is stopped
at a stationary X-point.  If the X-point is moving -- as surely it is
-- then stars can be squeezed across, much like icing from an icing
bag.  The relative number of stars trapped into libration about $\Pa$
or $\Pb$ is once more determined by the growth rates of the phase
space areas associated with the trapping region. When the pattern
speed alone is changing, the trapping region around $\Pb$ is
essentially excluding phase, whereas that around $\Pa$ is swallowing
phase.  Stationary X-points make the wave give up its bound angular
momentum.

\subsection{4.3 Figure Rotation and Triaxial Models}

\beginfigure{18}
\centerline{\psfig{figure=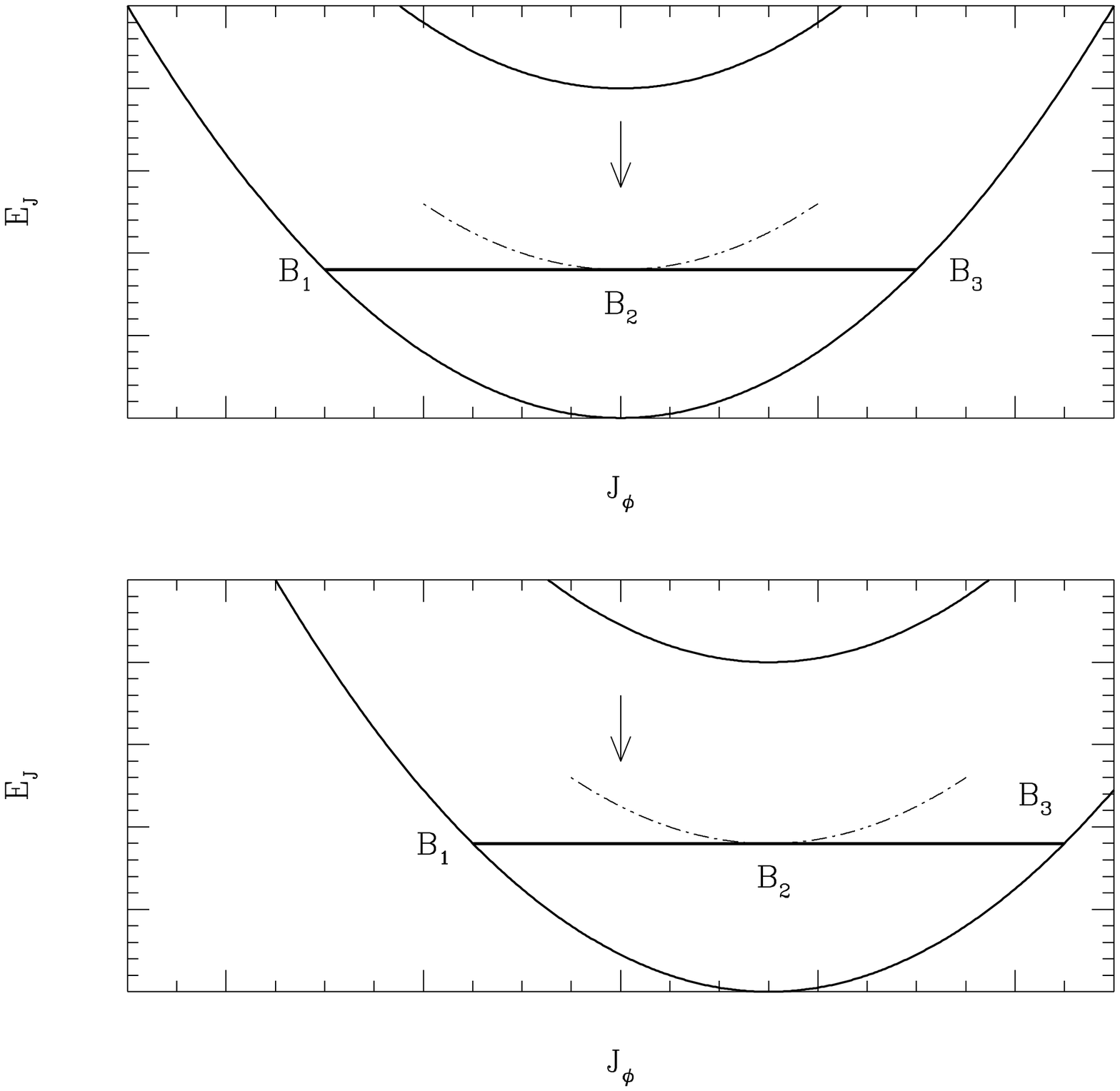,height=\hssize,width=\hssize}}

\caption{{\bf Figure 18.} The upper panel shows an equiaction section
at inner Lindblad resonance ($\ell =1, m = 2$). The bold horizontal
line represents a trapped box orbit in a galaxy with no figure
rotation. The lower bounding curve of the section represents the
trough of the well. So, $\Ba$ and $\Bc$ mark the points where the box
orbit crosses the trough. The dot-dashed line is the potential at the
slow angle $w$ marking the amplitude of libration of the box. This
occurs at $\Bb$, which coincides with $J_\phi =0$ (marked by an
arrow). So, the box orbit is bounded by straight line segments. The
lower panel shows the changes inflicted by rotation. The point $\Bb$
is now offset from $J_\phi =0$ and so the box orbit is no longer
bounded by straight lines.}
\endfigure

\beginfigure{19}

\centerline{\psfig{figure=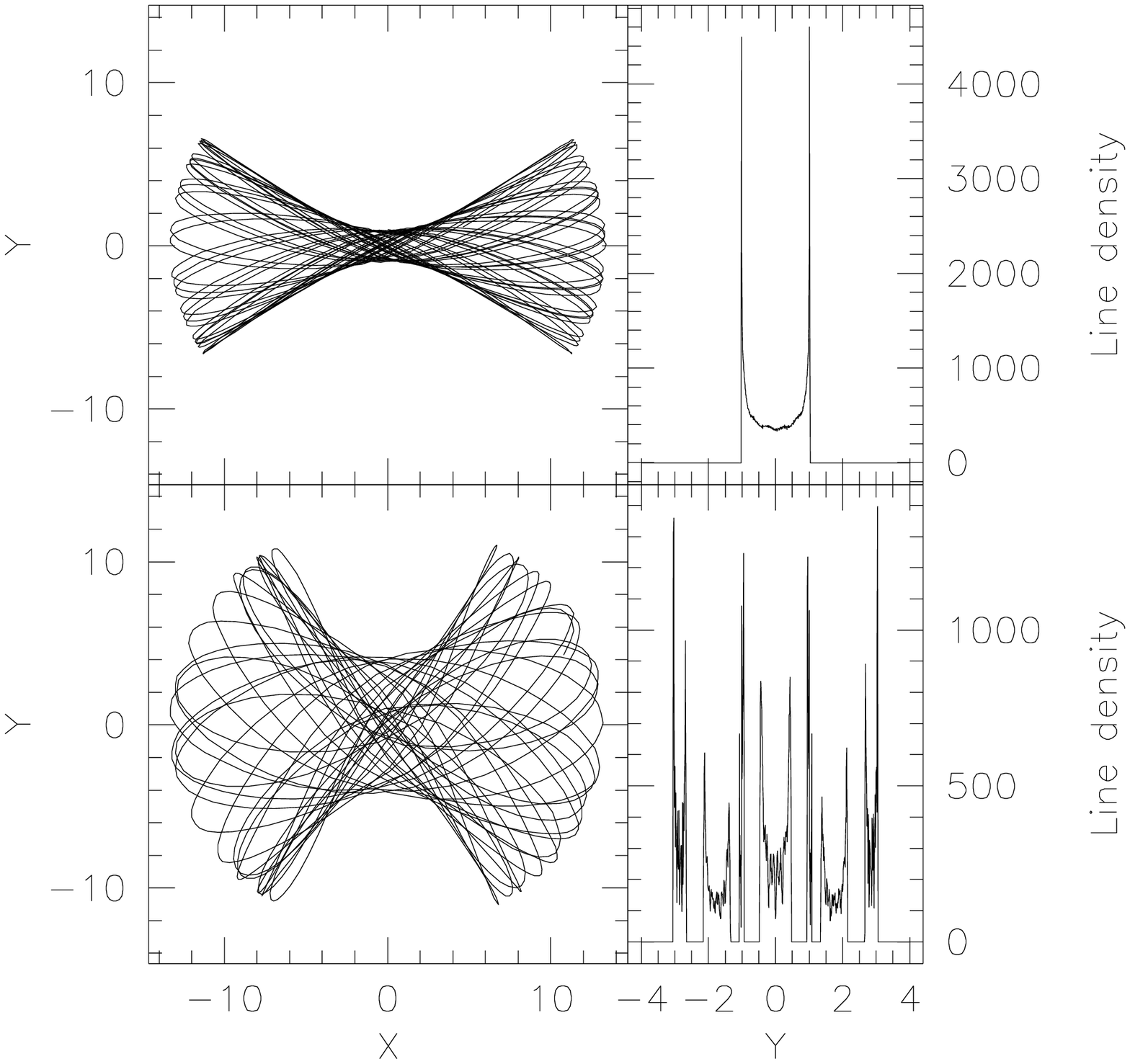,height=\hssize,width=\hssize}}

\caption{{\bf Figure 19.} The upper two panels show a box orbit in 
the principal plane of a stationary triaxial potential, together with
the orbital density across its waist. The two orbital density spikes
correspond to the bounding straight line sections of the box
orbit. The lower two panels show the same orbit in a triaxial
potential with figure rotation. The box orbit is bounded by looping
segments rather than straight lines. [The model used is the Binney
potential with $q=.9$ and $v_0=R_c=1$. The orbits are both launched
form the minor axis ($x=0, y=1$) with ($v_x = 2, v_y = -.2$). The
pattern speed $\Op =0$ in the former case and $0.015$ in the latter.]
}
\endfigure

\noindent

Analyses of images of galactic nuclei taken with the HST planetary
camera have shown that the surface brightness distributions of
early-type galaxies are almost always cusped (Lauer et al. 1996).  For
example, the surface brightness of the nearby S0 galaxy NGC 7547 is
cusped in the inner 600 parsecs like $R^{-1}$. Although there is no
difficulty in the sustenance of axisymmetric cusps (e.g., Evans 1994),
the survival of triaxial cusps is a much more delicate matter. For a
point perturber in a triaxial non-rotating galaxy, Gerhard \& Binney
(1985) computed the timescale on which a typical box orbit suffers a
serious deflection. They concluded that box orbits with apocentres
$\lta 1\,\kpc$ are disrupted over the course of a Hubble time. This
process may cause the shape of the inner parts of galaxies to become
axisymmetric. Schwarzschild and co-workers (Miralda-Escud\'e \&
Schwarzschild 1989; Lees \& Schwarzschild 1992) have also suggested
that strongly flattened triaxial figures with density cusps may not be
able to persist.  Recently, Merritt \& Fridman (1995) presented
numerical results on the existence of non-rotating triaxial galaxy
models with cusps. Only for weak central cusps ($\rho \sim r^{-1}$)
were they able to build equilibrium models, raising again the question
of the existence of stationary triaxial galaxies with strong central
cusps ($\rho \sim r^{-2}$). In this section, we suggest an answer to
the question : does figure rotation aid the survival of triaxial
models with central scatterers?

The upper and lower panels of Fig.~18 show equiaction sections in
static and mildly rotating triaxial potentials. A typical box orbit is
marked by a bold horizontal line in each case. B${}_1$ and B${}_3$
mark the points where the libration reaches the bottom of the well. At
B${}_2$, the libration finishes its swing. The dashed curve defines
the opening angle of the fan. When the potential is static, the box
orbit turns round in azimuth on the radial orbits ($J_\phi =0$). When
the potential is rotating, this turn-round occurs on an eccentric but
non-radial orbit ($J_\phi \neq 0$). The upper and lower panels of
Fig.~19 again refer to static and rotating triaxial potentials. In
each case, the orbit produced by propelling a star from the same spot
on the minor axis with the same speed is shown, together with the
linear density across the waist. In the non-rotating case, this
possesses cusps only at the waist's edge, but inner cusps appear when
there is figure rotation. The inner cusp is brought closer to the
centre by moving the point B${}_1$ in Fig.~18 towards $J_\phi =
0$. This subtlety aside, the more striking change is that the waist of
the box orbit has been broadened and the probability of hitting the
centre reduced. Waist broadening by figure rotation will therefore
assist the survival of triaxial galaxies with central scatterers.

We can also use the theoretical methods of this paper to estimate the
relative numbers of stars that are cast into the prograde and
retrograde directions when a bar dissolves. This may be relevant to
formation histories of galaxies like NGC 4550, which is built from two
similar counter--streaming stellar components (Evans \& Collett 1994).
A bar will consist of stars in a laminate of sections, and the capture
probability must be evaluated for each. In a strong bar, the
curvature of the envelope dominates over the central seam, but in the
final stages of dissolution or for weak bars, we can use the analysis
of Section 3.3. For a non-rotating bar, we lose equally into the two
streams. When there is rotation, the envelope of the section is
thinner towards the tip. Mild rotation is represented in the lower
panel of Fig.~18. The boxes are still very elongated and have a
positive inertial response, like rods. We can estimate the relative
size of the two streams as follows. The phase areas under the prograde
and retrograde parts of the lobe are:
\eqnam\twostream
$$\eqalign{S_{\rm prog} =&4 \Bigl| {A_2\over D_1} \Bigr|^{1/2}
+ \pi {3B_2D_1 - D_2A_2\over 3D_1^2},\cr
S_{\rm ret} =&4 \Bigl| {A_2\over D_1} \Bigr|^{1/2}
- \pi {3B_2D_1 - D_2A_2\over 3 D_1^2}.\cr}\eqno\new$$
A simple scale change of the envelope implies ${\dot B_2} =
{\dot A_2} B_2/A_2.$ The number of stars that escape into the two
streams is proportional to the changes in the two areas 
(\twostream). We find that the streams are weighted $\fr12 + \alpha :
\fr12 - \alpha$, with
\eqnam\niftyrule
$$\alpha = {\pi \over 12} {(3D_1 B_2 - D_2 A_2)\over
|D_1^3A_2|^{1/2}}.\eqno\new$$
Along an inner Lindblad section, the inertial response can change from
positive near the radial orbits to negative for near-circular orbits
(see the lower panel of Fig.~1). There is a point, then, at which
$D_1$ vanishes and the above analysis is not valid. For a weak, fast
rotating bar, we can return to the analysis of Section 3.2. In this
case, stars are more bound when the amplitude drops.  If the
rule-of-thumb (\niftyrule) is applied to the slow moving bar in the
upper panel of Fig.~4, then the number of stars moving in the prograde
sense is enhanced to $\sim 60$ per cent. This is by no means a small
asymmetry, especially in view of the slowness of the bar.  This
suggests that it is possible to build markedly asymmetric
counter-streams by break-up of a rapidly rotating bar.

\eqnumber =1
\def\chaphead{\hbox{5.}}
\section{5 Conclusions} 

This paper has shown how to calculate population changes caused by
resonant escape and capture in a disk of stars. These processes depend
on the shape of the effective potential well for orbital
capture. Changes in the well are easy to picture on an equiaction
section. For time-dependent problems, in particular, the equiaction 
section offers advantages over alternatives, such as Poincar\'e 
surfaces of sections. The main results of the paper are:
\medskip
\noindent
(1) There are barred galaxies with two outer rings of gas and stars
(so-called $\R1R2$ galaxies). It is very difficult to transfer stars
between the two outer rings in $\R1R2$ galaxies. Surprisingly, if the
bar is decelerating or dissolving, both the rings can grow. If the bar
is speeding up or increasing in strength, both the rings fade.
\medskip
\noindent
(2) Counter-rotating stars and gas are particularly susceptible to
large eccentricity change. This mechanism could be important in
channelling stars and gas towards the centres of galaxies. Tidal
resonant forcing of highly inclined orbits around a central massive
object will increase the likelihood of close encounters between the
orbiting star and the object. So, in the centres of galaxies, tidal 
forces -- perhaps caused by a sinking object -- can drive orbiting 
stars onto a black hole.
\medskip
\noindent
(3) Resonances can create sharp holes and notches in the stellar 
distribution function, as well as high velocity tails, the width
and shape of which we have explicitly calculated. The advection 
of angular momentum can be halted by the ocurrence of X-points
(defined in section 2.2).
\medskip
\noindent
(4) Figure rotation will assist the survival of triaxial, cusped
models by broadening the waists of box orbits. Moderate asymmetries
in the populations of prograde and retrograde stars are produced by 
dissolving mildy rotating bars. A counterstreaming disk with, say,
$60$ per cent of stars moving in the prograde, $40$ per cent in the
retrograde direction is a likely end-point of the disruption.

\section*{Acknowledgments}

It is a pleasure to thank Donald Lynden-Bell, Pak-Li Chau, Alar
Toomre, James Binney, David Earn and Christophe Pichon for useful
conversations. JLC acknowledges financial support from PPARC (grant
number GRJ 79454), while NWE is supported by the Royal Society.  David
Earn and Donald Lynden-Bell gave helpful and generous comments on the
draft version of the paper. We thank Scott Tremaine and Norm Murray
for drawing our attention to the work of Kozai and Bailey.

\eqnumber =1
\def\chaphead{\hbox{A}}
\section{Appendix A: The Jointed Arm}

\beginfigure{20}
%
\centerline{\psfig{figure=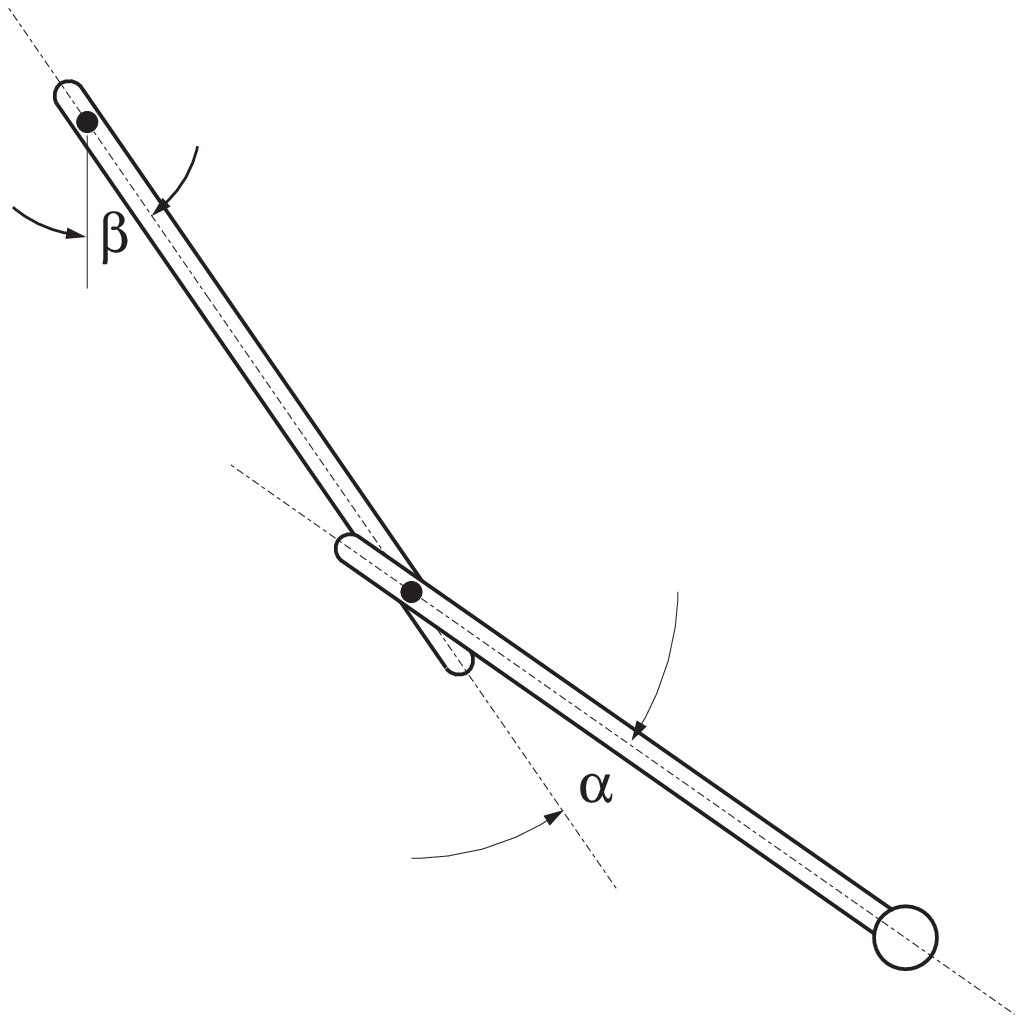,height=7.5truecm,width=7.5truecm}}
\caption{{\bf Figure 20.} A picture of the jointed arm.}
\endfigure

\beginfigure{21}
\centerline{\psfig{figure=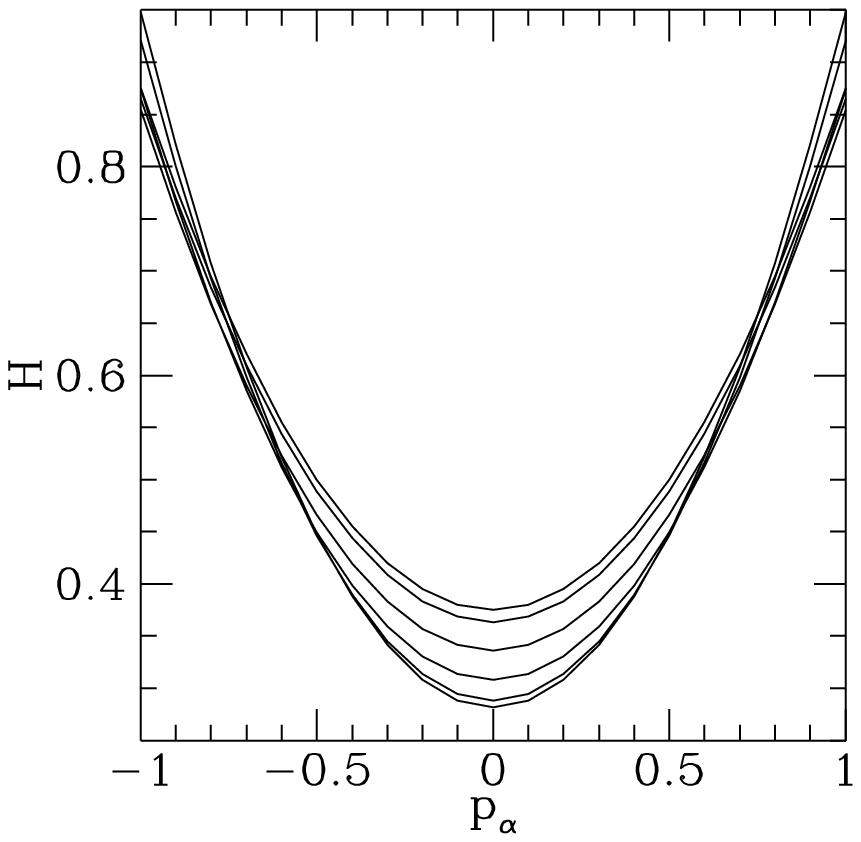,height=7truecm,width=7.5truecm}}
\caption{{\bf Figure 21.} A plot of the energy integral ($H$) against 
the canonical momentum $p_\alpha$ for the arm at a fixed value of the
canonical momentum $p_\beta$. This is analogous to the equiaction
sections discussed in the main body of the paper. The lines are drawn
at representative, equally spaced values of $\alpha$ between $0$ and
$\pi$. Trajectories on the section are horizontal lines bounded by the
envelope.}
\endfigure

\noindent
Fig.~20 depicts a simple mechanical system -- a mass attached to a
fixed pivot by a jointed arm. We suppose that all the inertia of the
arm resides in that part attached to the fixed pivot. The Lagrangian
of the system is just the kinetic energy, which we may write as
\eqnam\jarmlagrangian
$$L = (\mu +1){\dot \beta}^2 + {\dot \alpha}{\dot \beta} + {1\over 2}
{\dot \alpha}^2 + {\dot \beta}({\dot \alpha} + {\dot
\beta})\cos\alpha,\eqno\new$$
where $\mu$ measures the angular inertia of the arm relative to that
of the attached mass. We see that $\beta$ is a cyclic coordinate
manifesting the freedom to shift the azimuthal angle about the
pivot. The conserved momentum conjugate to $\beta$ is
$$p_\beta = {\partial L\over \partial {\dot \beta}} =
2\mu {\dot \beta} + (2{\dot \beta} + {\dot \alpha})(1+ \cos \alpha),
\eqno\new$$
which is of course the total angular momentum of the mass and the arm.
It is convenient to exploit this invariant to simplify the description
of motion in the remaining degree of freedom. To this end, we
construct the Routhian (Landau \& Lifshitz 1969),
$$R = p_{\beta} {\dot \beta} - L.\eqno\new$$
After dropping a total time derivative, which will not contribute to
the new action, we find
$$R = {p_\beta^2 - {\dot \alpha}^2(2\mu + \sin^2 \alpha)
\over 4(1 + \cos\alpha + \mu)}.\eqno\new$$
The equation of motion of $\alpha$, describing the pivoting of the
arm, is then obtained from the Routhian
$${d\over dt}\Bigl( {\partial R\over \partial {\dot \alpha}} \Bigr)=
{\partial R \over \partial \alpha},\eqno\new$$
or directly from an energy integral
$$\eqalign{H =& R - {\dot \alpha} {\partial R \over \partial
{\dot \alpha}},\cr
=& {p_\beta^2 + {\dot \alpha}^2(2\mu + \sin^2 \alpha)
\over 4(1 + \cos\alpha + \mu)}.\cr}\eqno\new$$
In the limit $\mu >>1$, when the inertia of the arm is much greater
than the attached mass, the librations reduce to that of the simple
pendulum
\eqnam\jarmeqofm
$${\ddot \alpha} = -{p_\beta^2 \over 4 \mu^2} \sin \alpha.\eqno\new$$
We can see that the effect of the conserved momentum is to provide an
effective potential for the swinging mass. The steady states of this
system (analogous to the periodic orbits) correspond to the arm
extended with $\alpha = 0$ (stable) or inwardly directed with
$\alpha = \pi$ (unstable).

We can represent the possible motions of the system on a section
through the phase space of the arm at fixed conserved momentum
$p_\beta$.  The ordinate in Fig.~21 is the energy integral
$$H = {p_\beta^2 \over 4(1 + \cos \alpha +\mu )} +
      {(1 + \cos \alpha + \mu)p_\alpha^2 \over 2\mu +
      \sin^2\alpha}.\eqno\new$$
This energy integral is conserved so the trajectories oscillate along
horizontal lines bounded by the confining envelope. These curves
correspond to the trough ($\alpha =0$) and the crest ($\alpha = \pi$)
of the potential, where the kinetic energy of the attached mass is at
a maximum or minimum. The greater the value of $p_\beta$, the greater
the breadth of the envelope and consequently the broader the libration
in $p_\alpha$. The curvature of the envelope is dictated by the
coefficient of $p_\alpha^2$. There are motions of the attached mass in
which it freely rotates about the joint pivot. There are motions too
in which the mass librates about an outward pointing radius
corresponding to the minimum of the effective centrifugal
potential. These trapped motions lie in the basin of the section.  In
this region, we see that the section is internally bi-symmetric and
dominated by a single harmonic.  The envelope constricts at large
$p_\alpha$.  There is an eventual cross-over of the lines of constant
$\alpha$ (c.f., Fig.~11).

We can illustrate too the r\^ole that adiabatic invariants can play in
the secular evolution of the system. Suppose the jointed arm, having
been uniformly heated, is slowly returning to its natural length.  We
can anticipate that the system will conserve angular momentum and spin
up as it shrinks. But, what happens to the libration of the attached
mass as a function of the length $\ell(t)$ of the arm? We have
implicitly removed a quadratic factor of length from the Lagrangian
(\jarmlagrangian) so that we can keep the equation of motion
(\jarmeqofm) but with respect to a scaled time $\tau$, such that
$$d\tau = {dt\over \ell (t)^2}.\eqno\new$$
As the arm shrinks, there is an effective increase in the centrifugal
acceleration, but the angular range of the oscillation remains the
same (as may easily be seen by integrating the equation of motion). In
this process, it may not -- for instance -- pass from rotation to
libration. Notice too that the action associated with the $\alpha$
oscillation is conserved exactly in this case. Suppose instead that
the inner arm though contracting overall at the same rate as the outer
has been heated differentially along its length. The re-distribution
of mass may lead $\mu$ to be a slowly varying function of time. In
this case, the action is only adiabatically invariant. Further, if
$\mu$ is decreasing (for example, if the inner arm experiences greater
heating close to the pivot), then the angular extent of the
oscillation also decreases. The existence of the adiabatic invariant
leads the mass to move closer to the minimum energy state in which the
arm is straight and aligned radially outward.

\eqnumber =1
\def\chaphead{\hbox{B}}
\section{Appendix B: Calculation of the Refined Capture 
Probability Formulae}

This Appendix gives some more details of the computations leading to 
the capture probabilities presented in Section 3.4. Suppose the 
Hamiltonian is 
$${\t H} = \Jt^2 - 2 \lambda \Jt - 2 (2 \Jt)^{1/2}
(1 - \epsilon \Jt) \cos \wt .\eqno\new$$
The separatrix actions $I$ are best evaluated using
\eqnam\goldreichs
$$I = {1\over \pi} \oint  \Jt d\wt = {1\over \pi} \oint {
\Jt{\dot \wt} \over {\dot \Jt} } d\Jt,\eqno\new$$
where the dots represent time derivatives and $\Jt (w)$ is the
equation of the separatrix. If the separatrix energy is $\Hsep$,
then the numerator becomes
$$\eqalign{ \Jt {\dot \wt} =& \fr32 [ 1 - \fr19 \epsilon (8\lambda -6\Jt
)](\Jt - \Jt_u)\cr
& \times [\Jt + \Jt_u - \fr23 \lambda - \fr{1}{27}\epsilon
(24\Hsep + 16\lambda^2)],\cr}\eqno\new$$
where $\Jt_u$ marks the position of the unstable fixed point.
The denominator can be factorised as
$${\dot \Jt} = [ (\Jt - \Jt_u)^2(\Jt_2 - \Jt)(\Jt - \Jt_1) ]^{1/2}.
\eqno\new$$
In other words, the unstable fixed point is always a double root,
whereas the ends of the separatrix $\Jt_1, \Jt_2$ are single roots
of the quartic. A lengthy calculation gives the areas under the 
separatrix branches as:
$$\eqalign{S_{\rm ret} &= \pi \lambda + 2\lambda \asin
                      ( x_\star^{-3/2}) + {3(x_\star^3 -1)^{1/2}
                      \over x_\star}\cr
             &-\epsilon [ {3\pi x_\star \over 2} 
                       + G_1(x_\star) ],\cr
S_{\rm prog} &= \pi \lambda - 2\lambda \asin ( x_\star^{-3/2} )
            - {3(x_\star^3 -1)^{1/2} \over x_\star} \cr
             &-\epsilon [{3\pi x_\star \over 2} 
                       - G_1(x_\star) ],\cr
S_{\rm trap} &= 4\lambda  \asin ( x_\star^{-3/2} ) + {6(x_\star^3
               -1)^{1/2} \over x_\star}.\cr 
             &-2\epsilon G_1(x_\star),}\eqno\new$$
where
$$G_1(x_\star) =  3\asin ( x_\star^{-3/2} ) + 
{(2 + x_\star^3)(x^3-1)^{1/2}\over x_\star^2}.\eqno\new$$
Using (\capturehenrard) now gives the final result for the capture 
probability reported in the text as (\jimsheroics).

The addition of a second harmonic at the U-point leads to
consideration of this Hamiltonian:
$${\t H} = {\t J}^2 - 2 \lambda {\t J} - 2 (2 {\t J})^{1/2}
\cos {\t w} - \mu {\t J} \cos 2{\t w},\eqno\new$$
The areas under the separatrices are:
$$\eqalign{S_{\rm ret} &= \pi \lambda + 2\lambda \asin
                      ( x_\star^{-3/2}) + {3(x_\star^3 -1)^{1/2}
                      \over x_\star}\cr
             &+\mu [ {\pi \over 2}G_2(x_\star) + G_3(x_\star)],\cr
S_{\rm prog} &= \pi \lambda - 2\lambda \asin ( x_\star^{-3/2} )
            - {3(x_\star^3 -1)^{1/2} \over x_\star} \cr
             &+\mu [  {\pi \over 2}G_2(x_\star)
                       - G_3(x_\star) ],\cr
S_{\rm trap} &= 4\lambda  \asin ( x_\star^{-3/2} ) + {6(x_\star^3
               -1)^{1/2} \over x_\star}.\cr
             &+2\mu G_3(x_\star),}\eqno\new$$
where
$$G_2(x_\star) = {1\over 36 x_\star^3 } \Bigl[x_\star^9 + 10 x_\star^6 -
14 x_\star^3 + 12 \Bigr].\eqno\new$$
$$G_3(x_\star) = (x_\star^3-1)^{1/2} \Bigl[ {x_\star^3\over 12} + {1\over
6x_\star^3} \Bigr] + G_2(x_\star) \asin ( x_\star^{-3/2} ).\eqno\new$$
In this problem, $x_\star \ge 1$ correspond to those values for 
which there is a trapping region. The capture probability is given
as (\jimsdeeds).  


\section*{References}

\beginrefs

\bibitem Allen C. W., 1973, Astrophysical Quantities. The Athlone
Press, London, chap. 7.

\bibitem Arnold V. I., 1989, Mathematical Methods of Classical 
Mechanics. Springer-Verlag, Berlin, second edition, chap. 10

\bibitem Arnold V. I., 1990, Huygens \& Barrow, Newton \& Hooke.
Birkh\"auser, Basel, p. 72

\bibitem Athanassoula E., Bosma A., Cr\'eze M., Schwarz M. P.,
1982, A\&A, 107, 101

\bibitem Athanassoula E., Bosma A., 1985, ARAA, 23, 147

\bibitem Bertola F., Buson L. M., Zeilinger W. W., 1992, ApJ, 401, L79

\bibitem Binney J. J., 1982, MNRAS, 201, 1

\bibitem Binney J. J., Tremaine S. D., 1987, Galactic Dynamics.
Princeton University Press, Princeton

\bibitem Binney J. J., 1987, in Gilmore G., Carswell B., The Galaxy.
Reidel, Dordrecht, p. 399

\bibitem Borderies N, Goldreich P., 1984, Celest. Mech., 32, 127

\bibitem Born M., 1927, The Mechanics of the Atom. G.Bell \& Sons,
London, chap. 4

\bibitem Braun R., Walterbos R., Kennicutt R., 1995, Nature, 360, 442

\bibitem Buta R., 1986, ApJS, 61, 609

\bibitem Cary J. R., Escande D. F., Tennyson J. L., 1986, Phys. Rev. A,
34,4256

\bibitem Ciri R., Bettoni D., Galletta G., 1995, Nature, 375, 661

\bibitem Cohen C. J., Hubbard E. C., 1965, AJ, 70, 10

\bibitem Collett J. L., 1995, Ph. D. thesis, Cambridge University,
chaps. 2, 3

\bibitem Contopoulos G., Papayannopoulos Th., 1980, A\&A, 92, 33




\bibitem Donner K. J., 1979, Ph. D. thesis, Cambridge University,
chap. 6

\bibitem Earn D. J. D., 1993, Ph. D. thesis, Cambridge University,
chap. 3

\bibitem Earn D. J. D., Lynden-Bell D., 1996, MNRAS, 278, 395

\bibitem Evans N. W., 1994, MNRAS, 267, 333

\bibitem Evans N. W., Collett J. L., 1994, ApJ, 420, L67

\bibitem Gerhard O. E., Binney J. J., 1985, MNRAS, 216, 467

\bibitem Goldreich P., 1965, MNRAS, 130, 159

\bibitem Goldreich P., Tremaine S. D., 1981, ApJ, 243, 1062
 
\bibitem Gutzwiller M. C., 1990, Chaos in Classical and Quantum Mechanics.
Springer-Verlag, Berlin, chap. 7

\bibitem Henrard J., 1982, Celest. Mech., 27, 3

\bibitem Henrard J., 1993, Dynamics Reported 2. Springer-Verlag,
Berlin, p. 117

\bibitem Henrard J., Lema\^itre A., 1983, Icarus, 55, 482

\bibitem Kalnajs A. J., 1973, Proc. Astron. Soc. Australia, 2, 174

\bibitem Kalnajs A. J., 1977, ApJ, 212, 637

\bibitem Kozai Y., 1980, Icarus, 41, 89

\bibitem Kuijken K., Tremaine S.D., 1991, in Sundelius, B., 
Dynamics of Disc Galaxies. G\"oteborg University Press, G\"oteborg, p. 71

\bibitem Landau L. D., Lifshitz E.M., 1969, Mechanics, Pergamon Press,
Oxford, p. 134

\bibitem Lauer T., et al., 1996, AJ, in press 

\bibitem Lees J., Schwarzschild M., ApJ, 384, 491

\bibitem Lidov M. L., 1963, in Problems of Motion of Artificial
Celestial Bodies. Akad. Nauk SSSR, Moscow, p 119 (in Russian)

\bibitem Lynden-Bell D., Kalnajs A. J., 1972, MNRAS, 157, 1

\bibitem Lynden-Bell D., 1973, in Dynamical Structure and Evolution
of Stellar Systems. Geneva Observatory, Sauverny, p. 91

\bibitem Malhotra R., 1993, Nat, 365, 819
 
\bibitem Merritt D., Fridman T., 1996, ApJ, 460, 136

\bibitem Mestel L., 1963, MNRAS, 126, 553

\bibitem Miralda-Escud\'e J., Schwarzschild M., 1989, ApJ, 339, 752

\bibitem Newton A., 1986, D. Phil. thesis, Oxford University

\bibitem Peale S. J., 1976, ARAA, 14, 215


\bibitem Sellwood J., Kahn F. D., 1991, MNRAS, 250, 278

\bibitem Sridhar S., Touma J., 1996, MNRAS, 279, 1273

\bibitem Stagg C., Bailey M., 1989, MNRAS, 241, 506

\bibitem Tremaine S. D., Weinberg M., 1984, MNRAS, 209, 729

\bibitem Toomre A., 1981, in  Fall S. M., Lynden-Bell D., The Structure 
and Evolution of Normal Galaxies. Cambridge University Press,
Cambridge, p. 111.

\bibitem Yoder C., 1973, Ph. D. thesis, University of California, Santa
Barbara

\bibitem Yoder C., 1979, Celest. Mech., 19, 3

\endrefs
\bye